\documentclass{elsart}
\newlength{\lslash}

\usepackage{amsmath}
\usepackage{amssymb,amsbsy,amsthm,amsopn,mathrsfs}
\usepackage{eufrak}
\usepackage{feynmp}
\usepackage{epsfig}
\usepackage{float}

\newcommand{\Lag}{\mathscr{L}}

\renewcommand{\vec}[1]{{\boldsymbol{#1}}}

\def\Im{\mathrm {Im}\,}
\begin{document}\setlength{\unitlength}{1mm}

\begin{frontmatter}
  
\title{Selfconsistent description of \\vector-mesons in matter}

\author{Felix Riek\thanksref{fr}} and \author{J\"o{}rn Knoll\thanksref{jk}}

\thanks[fr]{e-mail:f.riek@gsi.de}

\thanks[jk]{e-mail:j.knoll@gsi.de}

\address{Gesellschaft f\"u{}r Schwerionenforschung \\ Planckstr. 1 \\
  64291 Darmstadt}

\date{Feb. 2004}

\begin{abstract}
  We study the influence of the virtual pion cloud in nuclear matter
  at finite densities and temperatures on the structure of the $\rho$-
  and $\omega$-mesons. The in-matter spectral function of the pion is
  obtained within a selfconsistent scheme of coupled Dyson equations
  where the coupling to the nucleon and the $\Delta$(1232)-isobar
  resonance is taken into account. The selfenergies are determined
  using a two-particle irreducible (2PI) truncation scheme
  ($\Phi$-derivable approximation) supplemented by Migdal's short
  range correlations for the particle-hole excitations.  The so
  obtained spectral function of the pion is then used to calculate the
  in-medium changes of the vector-meson spectral functions. With increasing
  density and temperature a strong interplay of both vector-meson modes is
  observed. The four-transversality of the polarisation
  tensors of the vector-mesons is achieved by a projector technique.
  The resulting spectral functions of both vector-mesons and, through
  vector dominance, the implications of our results on the dilepton
  spectra are studied in their dependence on density and temperature.
\end{abstract}

\begin{keyword}

rho--meson, omega--meson, medium modifications, dilepton production,
self-consistent approximation schemes.

\PACS{14.40.-n}

\end{keyword}

\end{frontmatter}

\section{Introduction}
It is an interesting question how the behaviour of hadrons changes in
a dense hadronic medium. One of the driving ideas is that chiral
symmetry is expected to become partially restored with increasing
density or temperature. Therefore the spectral functions of hadrons in
their dependence on the thermodynamic parameters of the hadronic
environment are of special interest. They directly show the possible
effects on the mass and decay properties of the particles.  One of the
experimental accesses to observe the spectral functions in-matter is
provided through the study of electron-positron- or muon anti-muon
pairs, called dileptons \cite{HADES,BEVALAC,CERES,HELIOS,Ozawa} both
in hadron--nucleus and nucleus--nucleus collisions.  The advantage of
such electromagnetic probes instead of strong interacting particles,
like pions or kaons, is that one gets information directly from the
centre of the interaction zone.  Above invariant masses of 400 MeV the
pairs are mainly produced through the decay of vector-mesons like the
$\rho$- and $\omega$-meson. Additional contributions come from
bremsstrahlung and final-state Dalitz-decays of other resonances, like
the $\eta$-meson, which in part are controlled by other observations.
Presently all nuclear collision experiments
\cite{BEVALAC,CERES,HELIOS,Ozawa} observe a significantly enhanced
dilepton yield in the invariant mass range below the light
vector-meson masses compared to straight extrapolations from elementary
processes, like proton-proton scattering.  Several mechanisms,
especially a lowering of the $\rho$-meson mass or an increase of its
damping width, were proposed to explain the experimental facts
\cite{Peters1,Post1,Rapp,Urban,Brown,Kondratyuk,Lutz3,Herrmann,Rapp2,%
  Eletsky,Haglin,Pisarski,Wolf,Roy,Kaiser,Kaempfer,Hatsuda,Leupold,Jin}.
Additional experiments for photo-production off nuclei are planned,
where one likes to investigate the
$\omega\rightarrow\pi^0\gamma\rightarrow 3\gamma$ channel
\cite{Messch,Metag}. Despite the perturbation through the final state
interaction of the $\pi^0$-meson \cite{Messch} it will still be possible to
isolate the $\omega$-meson component because the competing
$\rho$-meson process is highly suppressed.

Ever since the early considerations of vector-mesons in dense matter
\cite{GaleKapusta,KorpaPratt,Herrmann} much effort has been devoted to
study the properties of the $\rho$-meson, which strongly couples to
the two-pion channel. However, with a few exceptions, e.g.
\cite{Rapp}, the in-medium properties of the pion are commonly
neglected, just employing unperturbed free pion states.  Furthermore
so far there are only few attempts to study the in-medium properties
of the $\omega$-meson
\cite{Eletsky,Haglin,Pisarski,Wolf,Roy,Kaiser,Schneider1,Saito,Wachs}.
Thus, a more realistic determination of the in-matter pion cloud
permits to investigate the dependence of the vector-meson properties
on the various components of the pion spectral function in the medium.
Compared to a schematic three level model for the pion modes studied
by Wachs \cite{Wachs} we use a scheme of coupled Dyson equations in
which the pionic degrees of freedom are modified selfconsistently
through the coupling to the nucleon and the $\Delta$(1232)-resonance.
All selfenergies are determined within a truncated two-particle
irreducible (2PI) effective action formalism ($\Phi$-functional) in
their full dependence on energy and three momentum.  This guarantees
conservation laws of the theory to be fulfilled on the level of
expectation values \cite{Baym1} despite the approximations employed.
From the in-medium spectral function of the pion we are able to
calculate the modified spectral functions for the $\rho$-meson as well
as for the $\omega$-meson in coupled scheme. Thereby the process
$\rho\rightarrow\omega +\pi$ leads to a strong interplay between both
vector-mesons.  In particular the $\omega$-width is very sensitive to
the space-like, i.e. low energy component of the pion caused by its
coupling to nucleon nucleon-hole states.  For a realistic treatment of
this component Migdal's short range correlations \cite{migdal} are
important.  These RPA-type correlations shift the spectral strength to
higher energies, such that pion condensation is prevented. In order to
concentrate on the effects due to the modified pion we neglected
direct couplings of the vector-mesons to the baryonic currents. The
latter have been investigated by several authors \cite{Peters1,Post1},
who found a multicomponent structure for the in-medium $\rho$ meson
which is yet absent in our approach.

Dealing with vector-mesons requires to consider current
conservation on the correlator level. This is ensured by a projector
formalism \cite{Hees1} which constructs a four-transversal
polarisation tensor.
        
The paper is organised in the following way: We first summarise the
essential definitions in the realtime formalism used in the
calculations at finite temperature and density.  Then we present the
model for the in-medium pion coupled to the the baryonic degrees of
freedom.  Afterwards we show how to derive a four-transversal and thus
a current conserving polarisation tensor by means of a projector
formalism.  For the so obtained in-medium spectral functions of the
vector-mesons we use the vector-dominance concept in order to
calculate the effect of our results on the dilepton rates. The
different decay contributions as well as the effect of possible
mass-shifts of the vector-mesons in the dilepton spectra are
investigated. A discussion of our results concludes the paper.

\section{Realtime formalism and 2PI generating functional}

Interested in all spectral properties as a function of real energies
the realtime formalism \cite{Kad-Baym} will be used rather than the
Matsubara technique.  Within this formalism all two-point functions
such as the propagator\footnote{We use units with $c=\hbar =1$ throughout the
paper.} or selfenergies take four components, e.g.
\begin{eqnarray}
        iG^{--}(x,x^\prime)
        &\equiv \langle T\hat{\Psi}(x)\hat{\Psi}^\dagger(x^\prime)\rangle
        \nonumber\quad
        iG^{++}(x,x^\prime)
        &\equiv \langle T^{-1}\hat{\Psi}(x)\hat{\Psi}^\dagger(x^\prime)\rangle
        \nonumber\\
        iG^{+-}(x,x^\prime)
        &\equiv \langle\hat{\Psi}(x)\hat{\Psi}^\dagger(x^\prime)\rangle
        \nonumber\quad\;\;
        iG^{-+}(x,x^\prime)
        &\equiv \mp\langle\hat{\Psi}^\dagger(x^\prime)\hat{\Psi}(x)\rangle,
\end{eqnarray}
depending on the order of the fermionic (upper sign) or bosonic (lower
sign) field operators $\hat{\Psi}(x)$ appearing in the expectation
values. The space-time coordinates are denoted by $x$ and $x'$, the
time-, respectively, anti-time-ordering is symolized by $T$ and
$T^{-1}$.  The Dyson equations then become matrix equations with
respect to the \{$-,+$\} labels (placement on the closed time
contour).

For a uniform medium in thermal equilibrium one can exploit the
standard simplification by going to the space-time
Fourier-transformation $x-x^\prime\rightarrow p$ of all quantities,
where $p=(p^0,{\vec{p}})$ is the four momentum.  Then the coupled
Dyson equations simply become algebraic and it is sufficient to solve
for the retarded (or advanced) Dyson equation, since all four
components can be deduced from this by thermal Fermi-Dirac or
Bose-Einstein factors arising from the Kubo-Martin-Schwinger (KMS)
condition.  Thereby the retarded and advanced Green functions follow
from
\begin{equation}\label{retard}
\begin{split}
        G^R(p)&\equiv G^{--}(p)-G^{-+}(p)=G^{+-}(p)-G^{++}(p)\\
        G^A(p)&\equiv G^{--}(p)-G^{+-}(p)=G^{-+}(p)-G^{++}(p)\\
        G^A(p)&\equiv(G^R(p))^\dagger
\end{split}
\end{equation}
with the KMS conditions
\begin{equation}
\begin{split} \label{KMS}
        iG^{-+}&\equiv \mp n_T(p)A(p)\\
        iG^{+-}&\equiv (1\mp n_T(p))A(p),
\end{split}
\end{equation}
where the spectral function and thermal weights are given by
\begin{equation}
\begin{split}
        iA(p)&\equiv G^A(p)- G^R(p)= (G^R(p))^\dagger- G^R(p)\\
        n_T(p)&\equiv\frac{1}{\exp((p^0-\mu)/T)\pm 1}
\end{split}
\end{equation}
with temperature $T$, chemical potential $\mu$ and the energy $p^0$.
The above listed relations for $G$ likewise apply for all equilibrium
two-point functions, i.e. also for the selfenergies. They are also
written in a way that they apply to Dirac spinors or vector fields,
whereby all quantities additionally become Dirac-matrix or
Lorentz-tensor structures, respectively. The adjungation denoted by
the $\dagger$-symbol refers to this matrix structure\footnote{ Dirac
  spinors need to be treated with special care since one has to employ
  the ``hermitian'' form $\gamma^0\gamma^\mu$ of the Gamma-matrices
  before executing the adjungation.}.  For relativistic bosons the
spectral functions $A(p)$ are normalised to $2\pi$ with respect to the
integration measure $p^0\; dp^0$.

The propagators and selfenergies of the baryons are denoted by $G$ and
$\Sigma$, while we use $D$ and $\Pi$ for the corresponding quantities
for the meson fields. In selfconsistent two-particle irreducible (2PI)
approximation schemes \cite{Baym1,Ivanov} the selfenergies $\Sigma$ or
$\Pi$ for the baryons and mesons are derived from a generating
functional, called $\Phi$-functional. This functional is given by a
truncated set of closed diagrams in accordance with the interaction
Lagrangian where all lines denote dressed, i.e. selfconsistent
propagators. The selfenergies result as functional variations with
respect to the propagators, i.e.
\begin{equation}
        -i\Sigma(x,y)=\frac{\delta i\Phi[G,D]}{\delta iG(y,x)}
        \qquad\mbox{or}\qquad
        -i\Pi(x,y)=\frac{\delta i\Phi[G,D]}{\delta iD(y,x)}.  
        \label{selfenergy}
\end{equation}
This implies an opening of a corresponding propagator line in the
diagrams of $\Phi$. For the resulting set of coupled Dyson equations
such 2PI approaches guarantee that even in a partial resummation of a
single class of diagrams the conservation laws which are related to
the symmetries of the system are fulfilled on the level of expectation
values \cite{Baym1}.

\section{The Model}

The present investigations are restricted to spin as well as isospin
symmetric nuclear matter. This implies vanishing chemical potentials
for all mesons in the light flavour sector. All selfconsistent
calculations of the spectral functions are carried out numerically as
a function of energy and momentum on a two dimensional grid of 5 MeV
by 5 MeV/c resolution using an iterative procedure.

\subsection{Pions and baryonic Resonances}\label{subsect-piNDelta}
In the baryon sector the nucleon and the $\Delta$(1232)-resonance are
included. The pion is considered as Goldstone boson of chiral symmetry
which, in lowest approximation in pion-energy, leads to $p$-wave
couplings to the baryonic pseudo-vector currents together with a minor
$s$-wave term which we ignore. For the energies involved we approximate
the spinor structure of the baryons by the non-relativistic limit,
while the kinematics is kept in the relativistic form. This
approximation allows to handle the spin-3/2-structure of the
$\Delta$-resonance in a simple way\footnote{ A fully relativistic
  formulation would require the use of special interactions as can be
  seen for example in ref. \cite{Pascalutsa}, which then would also
  have to be incorporated in the later introduced Migdal formalism.}.
Furthermore, for the investigated range in excitation energies we can
safely use the no-sea approximation for the baryons. Thus we neglect
processes involving anti-baryons and set the baryon spectral functions
to zero for negative energies.

The following non-relativistic form of the interaction Lagrangian
\begin{equation}
         {\Lag}^{\rm{int}}_{\rm{N}\Delta\pi}= -
         g_{\pi N N}{\Psi^\dagger}_N\left(\vec{\sigma}^\dagger\vec{\nabla}
         \right)\left(\vec{\tau}^\dagger\vec{\pi}\right)\Psi_N
         -g_{\pi N
         \Delta}{\Psi^\dagger}_\Delta\left(\vec{S}^\dagger\vec{\nabla} 
         \right)\left(\vec{T}^\dagger\vec{\pi}\right)\Psi_N
         +h.c.\label{Baryon-lag} 
\end{equation}
is used with the spinor-field operators $\Psi_N$ and $\Psi_\Delta$ of
the nucleon and $\Delta$-resonance. Here $\vec{\sigma}$ and
$\vec{\tau}$ denote the vectors of Pauli matrices in spin and isospin,
respectively. Likewise, the vector $\vec{S}$ defines the three
spin-coupling matrices of spin 3/2 and 1/2 to spin 1, while $\vec{T}$
is the same in isospin-space. In our scheme we neglect pion
self-interactions (${\Lag}^{\rm{int}}_{\pi}\propto \pi^4$) which give
rise to selfenergy diagrams of sunset type \cite{Hees1} as well as
contributions from $\Delta\Delta^{-1}$ excitations. Both are of some
importance at temperatures above 80 MeV, where they would even further
broaden the pion spectral functions.

Furthermore we supplement an exponential formfactor
\begin{equation}
        F(\textbf{q})=\exp(-\textbf{q}^2/\Lambda^2)    \label{formfactor}
\end{equation}
at the $\pi$NN- and $\pi$N$\Delta$-vertices. Values of $\Lambda$= 440
MeV and $g_{\pi N \Delta}=0.02\; \mbox{MeV}^{-1}$ are required for a
decent fit of the pion-nucleon phase-shifts in the
33-chanal\footnote{This implies a cut-off dependence of the in-medium
  results. In models with a larger body of resonances, cf.
  \cite{DmitrievSuzuki}, the cut-off $\Lambda$ can be chosen above 700
  MeV.} and a proper vacuum spectral function of the
$\Delta$-resonance.  For $g_{\pi N N}$ we use 0.007 $\mbox{MeV}^{-1}$
\cite{Weinhold}.

For the chosen Lagrangian mean-field terms drop out and the first
diagrams for the $\Phi$-functional in the baryon-pion
sector which we use are:
\begin{fmffile}{PHIBaryon}
\begin{displaymath}
\Phi_{N\Delta\pi}=
\parbox{40\unitlength}{
\begin{fmfgraph*}(40,25)
        \fmfpen{thick}
        \fmfleft{i}
        \fmfright{o}
        \fmf{phantom}{i,v1}
        \fmf{phantom}{v2,o}
        \fmf{plain_arrow,left=1,tension=.2,label=$N$}{v1,v2}
        \fmf{plain_arrow,left=1,tension=.2,label=$N$}{v2,v1}
        \fmf{dashes,tension=.2,label=$\pi$}{v1,v2}
\end{fmfgraph*}}
+
\parbox{40\unitlength}{
\begin{fmfgraph*}(40,25)
        \fmfpen{thick}
        \fmfleft{i}
        \fmfright{o}
        \fmf{phantom}{i,v1}
        \fmf{phantom}{v2,o}
        \fmf{plain_arrow,left=1,tension=.2,label=$N$}{v2,v1}
        \fmf{dbl_plain_arrow,left=1,tension=.2,label=$\Delta$}{v1,v2}
        \fmf{dashes,tension=.2,label=$\pi$}{v1,v2}
\end{fmfgraph*}}
\end{displaymath}
\end{fmffile}%
The lines represent the fully dressed propagators, while apart from
the formfactor (\ref{formfactor}) the vertices remain bare.

Not all of the selfenergy terms obtained through (\ref{selfenergy})
are of equal importance. For instance the $\Delta$-$\pi$-loop
contribution to the nucleon selfenergy is found negligible compared to
the $N$-$\pi$-loop and thus neglected in the production runs. The
retarded pion selfenergy can be obtained from a dispersion relation
which in no-sea approximation does not require any renormalization
\begin{equation}
\begin{split}
        \Pi^R(p)=\frac{i}{2\pi}\int dE^\prime
        \frac{\Pi^{+-}(E^\prime,\vec{p})-\Pi^{-+}(E^\prime,\vec{p})}
        {E^\prime-E-i\epsilon} \label{SigmaR}.
\end{split}
\end{equation}
Interested in the pion modes a complete determination of the baryon
selfenergies $\Sigma(p)$ is beyond the scope and not the aim of the
present model.  For simplicity we drop the real parts of $\Sigma(p)$,
which would require a renormalization procedure within a
selfconsistent resummation scheme \cite{vanHeesKnoll}. However the
normalisation of the corresponding spectral functions is restored in
each iteration step.

\begin{figure}[H]
        \begin{minipage}[t]{.48\linewidth}
         \epsfig{file=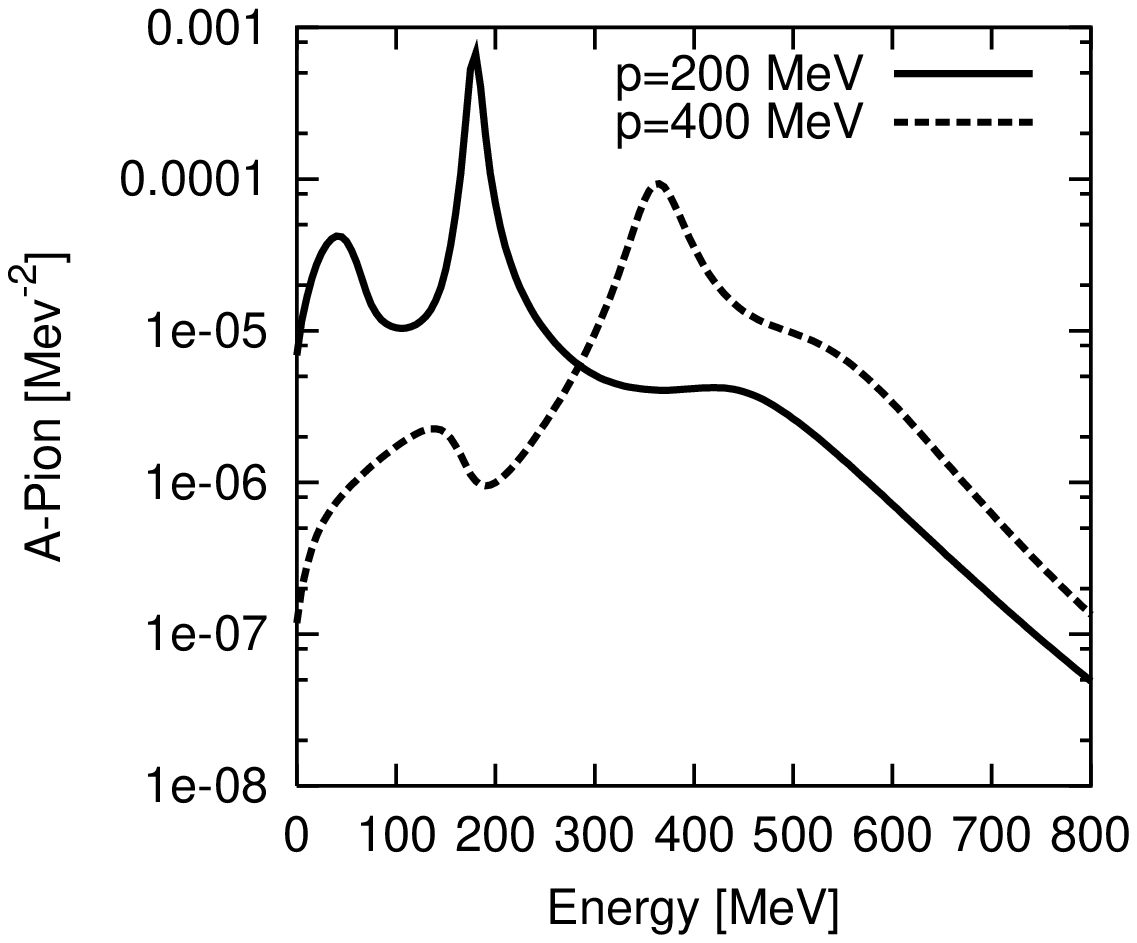,height=6.6 cm,width=5.41 cm,angle=-90}
         \caption{Pion spectral function $ A_\pi$ at T=0 MeV and $\rho
           =\rho_0$ for two different momenta\label{fig1}}
         \end{minipage}\hfill
        \begin{minipage}[t]{.48\linewidth}
         \epsfig{file=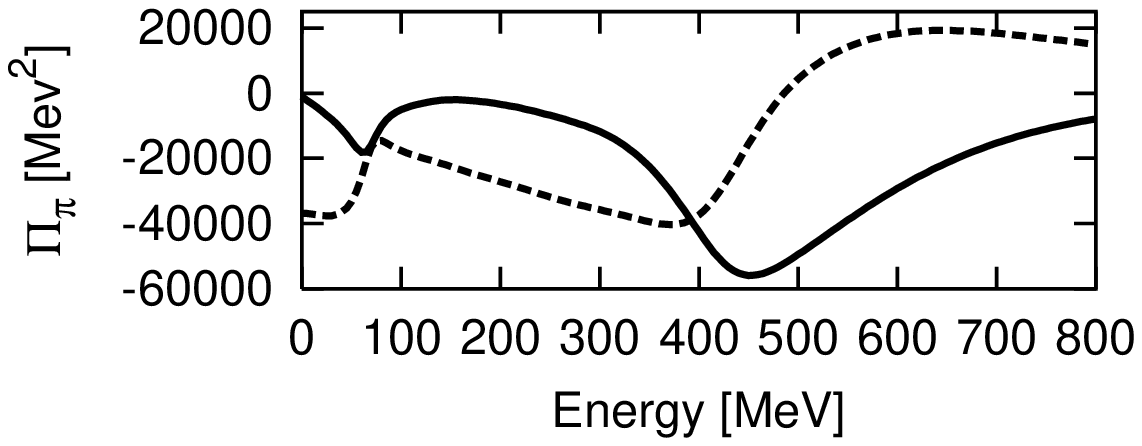,height=6.6 cm,width=2.7 cm,angle=-90}
         \epsfig{file=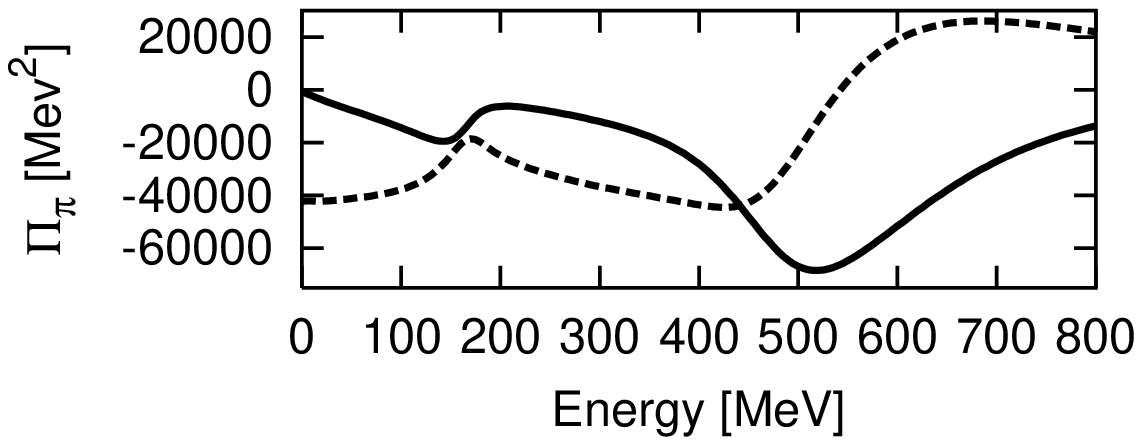,height=6.6 cm,width=2.7 cm,angle=-90}
         \caption{Real (dotted  line) and imaginary (full line)
                  part of the pion selfenergy at T=0 MeV
                  and $\rho =\rho_0$ for $|\textbf{p}|=200$ MeV (upper panel) and
                  $|\textbf{p}|=$ 400 MeV (lower panel)\label{fig2}}
         \end{minipage}
         \end{figure}

\begin{figure}[H]
        \begin{minipage}[t]{.48\linewidth}
         \epsfig{file=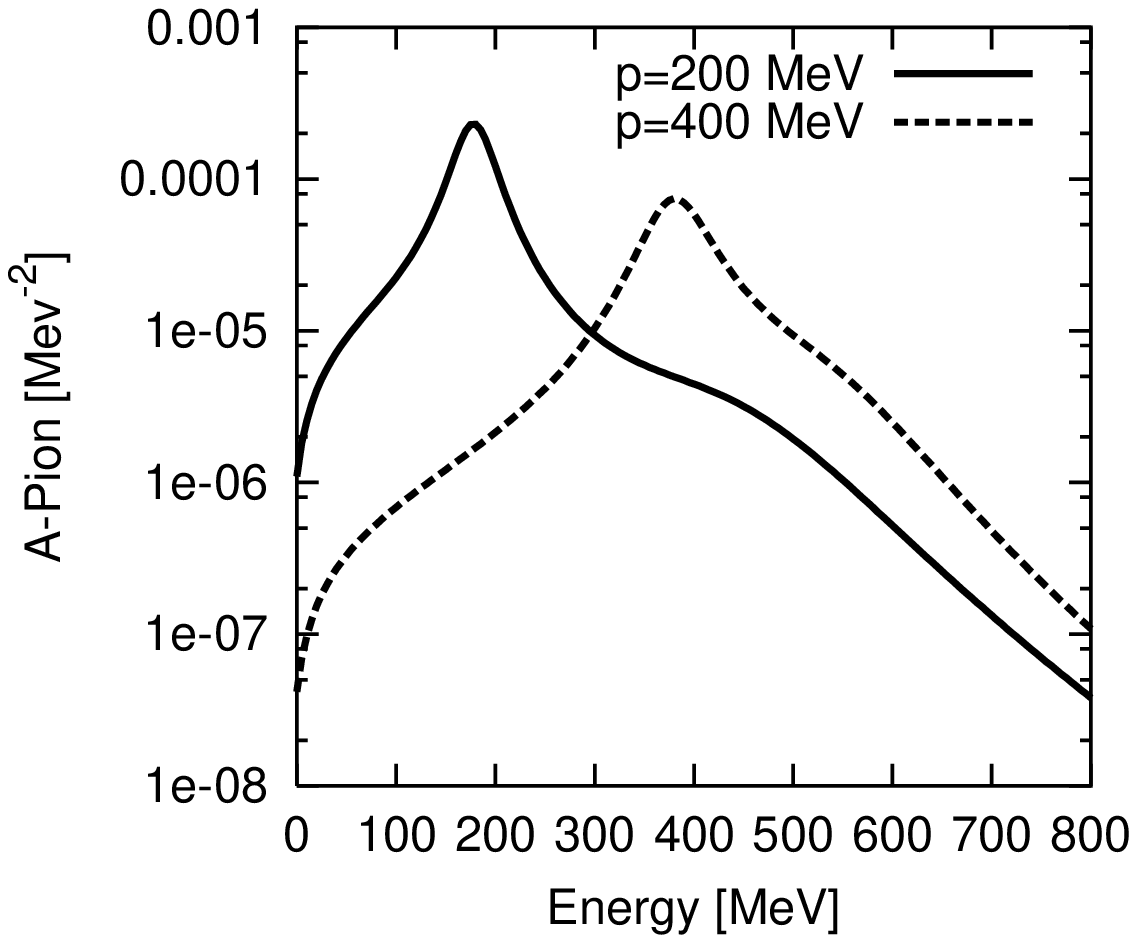,height=6.6 cm,width=5.41 cm,angle=-90}
         \caption{Same as Fig. \ref{fig1} at T=120 MeV \label{pion-spec}}
         \end{minipage}\hfill
        \begin{minipage}[t]{.48\linewidth}
         \epsfig{file=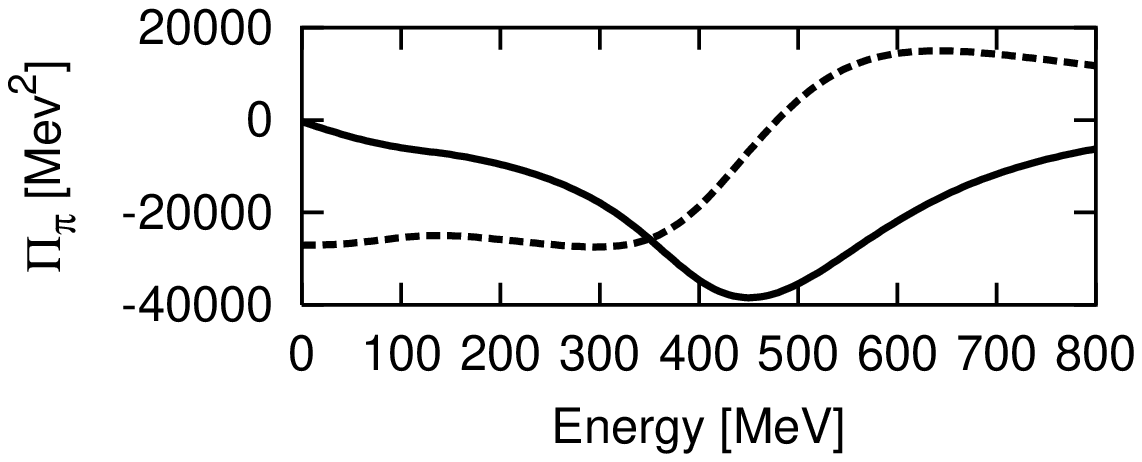,height=6.6
         cm,width=2.7 cm,angle=-90} 
         \epsfig{file=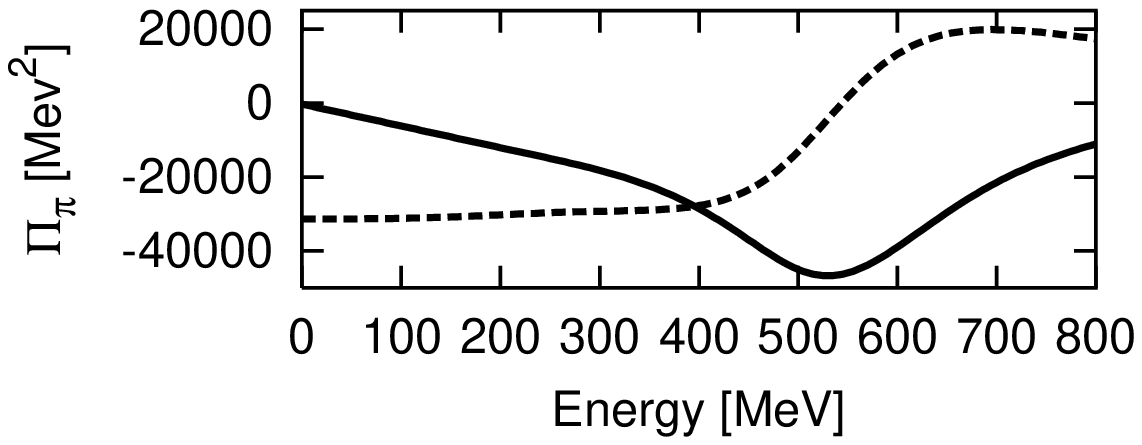,height=6.6
         cm,width=2.7 cm,angle=-90} 
         \caption{Same as Fig. \ref{fig2} at T=120 MeV\label{fig4}}
         \end{minipage}
\end{figure}

As is well known, the coupling of the pion to nucleon-hole ($NN^{-1}$) and
$\Delta N^{-1}$-states provides a strong softening of the pion modes already
at normal nuclear densities. Since there are no indications from
experiment for a pion condensate at normal nuclear density one has to
include a repulsive force which shifts the spectral strength up to
higher energies. This can be achieved by a repulsive short-range
interaction between the baryons first introduced by Migdal
\cite{migdal}
\begin{center}
\begin{fmffile}{Migdalvertex1}
\begin{fmfgraph*}(25,12)
        \fmfpen{thick}
        \fmfleft{i,i2}
        \fmfright{o,o2}
        \fmf{plain_arrow}{i,v1}
        \fmf{plain_arrow}{v1,o}
        \fmf{plain_arrow}{v1,i2}
        \fmf{plain_arrow}{o2,v1}
        \fmfv{l=$g_{11}$,l.d=15}{v1}
\end{fmfgraph*}
\end{fmffile}
\begin{fmffile}{Migdalvertex2}
\begin{fmfgraph*}(25,12)
        \fmfpen{thick}
        \fmfleft{i,i2}
        \fmfright{o,o2}
        \fmf{plain_arrow}{i,v1}
        \fmf{plain_arrow}{v1,o}
        \fmf{dbl_plain_arrow}{v1,i2}
        \fmf{plain_arrow}{o2,v1}
        \fmfv{l=$g_{12}$,l.d=15}{v1}
\end{fmfgraph*}
\end{fmffile}
\begin{fmffile}{Migdalvertex3}
\begin{fmfgraph*}(25,12)
        \fmfpen{thick}
        \fmfleft{i,i2}
        \fmfright{o,o2}
        \fmf{plain_arrow}{i,v1}
        \fmf{plain_arrow}{v1,o}
        \fmf{dbl_plain_arrow}{v1,i2}
        \fmf{dbl_plain_arrow}{o2,v1}
        \fmfv{l=$g_{22}$,l.d=15}{v1}
\end{fmfgraph*}
\end{fmffile}
\end{center}
For the Migdal parameters we used $g_{11}=0.5$, $g_{22}=0.6$
$g_{12}=0.5$ without any form factor. At a momentum scale of 200 MeV/c
for the relevant $NN^{-1}$ excitations\footnote{These excitations,
  which essentially influence the $\rho$-$\omega$ mixing in matter, are
  mainly determined by $g_{11}$.} this choice is compatible with the
recent parameter set of Suzuki \cite{Suzuki} and Nakano \cite{Nakano}
who calibrated it on Gamow-Teller transition strengths.  These
correlations are to be included in the retarded pion selfenergy in
form of an RPA-resummation with particle-hole loops
\begin{fmffile}{Repulsion}
\begin{displaymath}
\Pi^R_\pi=
\parbox{40\unitlength}{
\begin{fmfgraph*}(40,15)
        \fmfpen{thick}
        \fmfleft{i}
        \fmfright{o}
        \fmf{dashes}{i,v1}
        \fmf{dashes}{v3,o}
        \fmf{plain_arrow,left=1,tension=.5}{v2,v1}
        \fmf{plain_arrow,left=1,tension=.5}{v1,v2}
        \fmf{plain_arrow,left=1,tension=.5}{v2,v3}
        \fmf{plain_arrow,left=1,tension=.5}{v3,v2}
\end{fmfgraph*}}
\;+\;
\parbox{40\unitlength}{
\begin{fmfgraph*}(40,15)
        \fmfpen{thick}
        \fmfleft{i}
        \fmfright{o}
        \fmf{dashes}{i,v1}
        \fmf{dashes}{v4,o}
        \fmf{plain_arrow,left=1,tension=.4}{v2,v1}
        \fmf{plain_arrow,left=1,tension=.4}{v1,v2}
        \fmf{dbl_plain_arrow,left=1,tension=.4}{v2,v3}
        \fmf{plain_arrow,left=1,tension=.4}{v3,v2}
        \fmf{plain_arrow,left=1,tension=.4}{v3,v4}
        \fmf{plain_arrow,left=1,tension=.4}{v4,v3}
\end{fmfgraph*}}
\;+\;\dots,
\end{displaymath}
\end{fmffile}%
where also mixtures with the $\Delta$-resonance, i.e $\Delta N^{-1}$-loops
need to be considered.  In the non-relativistic limit for the spinor
structure of the baryons, only the spatial components of the Dirac
matrices survive in all loops which, by symmetry arguments, are diagonal
(for a fully relativistic treatment cf.  \cite{Lutz2}). Therefore,
apart from a $\textbf{q}^2$-factor due to the difference between the
pion-baryon coupling and the Migdal interaction the basic loops for
the RPA-resummation are given by the normal pion selfenergy diagrams
resulting from the $\Phi$-functional (\ref{selfenergy})
\begin{fmffile}{SelfNuc}
\begin{displaymath}
\Pi^R_{\pi\;NN^{-1}}=
\parbox{40\unitlength}{
\begin{fmfgraph*}(25,15)
        \fmfpen{thick}
        \fmfleft{i}
        \fmfright{o}
        \fmf{dashes}{i,v1}
        \fmf{dashes}{v2,o}
        \fmf{plain_arrow,left=1,tension=0.4}{v1,v2,v1}\fmffreeze
\end{fmfgraph*}}
\Pi^R_{\pi\;\Delta h}=
\parbox{40\unitlength}{
\begin{fmfgraph*}(25,15)
        \fmfpen{thick}
        \fmfleft{i}
        \fmfright{o}
        \fmf{dashes}{i,v1}
        \fmf{dashes}{v2,o}
        \fmf{plain_arrow,left=1,tension=.4}{v2,v1}
        \fmf{dbl_plain_arrow,left=1,tension=.4}{v1,v2}
\end{fmfgraph*}}
\end{displaymath}
\end{fmffile}
Thus, the RPA-resummation can be done without further problems. With
the three possible vertices one finally obtains \cite{DmitrievSuzuki,Urban} 
\begin{equation}
        \Pi^R=(\textbf{q}^2)^2\frac{\Pi^R_{\pi\; NN^{-1}}
          +\Pi^R_{\pi\; \Delta h}-(g_{11}-2g_{12}
          +g_{22})\Pi^R_{\pi\;NN^{-1}}\Pi^R_{\pi\;\Delta h}}
        {(\textbf{q}^2-g_{11}\Pi^R_{\pi\;NN^{-1}})(\textbf{q}^2
          -g_{22}\Pi^R_{\pi\;\Delta h})
          -g_{12}^2\Pi^R_{\pi\;NN^{-1}}\Pi^R_{\pi\;\Delta h}}F(\textbf{q})
        \label{Migdal1}
\end{equation}
for the retarded pion selfenergy. The corresponding results together
with those for the pion spectral function are shown in the Figs.
\ref{fig1} to \ref{fig4}.
In the selfenergy and spectral function we see the effect of the
coupling to the different baryonic excitations. The on-shell pole
broadens and is shifted towards lower energies due to the
net-attraction of the interaction in this kinematical region. This
component strongly mixes with the $\Delta N^{-1}$ component which
leads to the shoulders around this peak. This component appears in the
time-like region for low momenta, while it traverses to the space-like
region for higher momenta. At low energies an entirely space-like pion
component appears due to $NN^{-1}$ excitations. From the kinematics it
is clear that this component stays in the space-like region for all
momenta.  Such space-like pion modes have to be interpreted as a
scattering process on the baryons in the matter mediated by the
exchange of pions. Due to this component, the pion spectral function
has non-vanishing strength at all energies, such that in all processes
where pions are involved all thresholds disappear.  Due to the
short-range Migdal repulsion there is no spectral component where the
peak position decreases towards lower energies when one increases the
momentum.  The form of the spectral function turns out to be very
sensitive on values for the Migdal parameters and the formfactor $F$.
In addition, in the literature one finds quite different procedures to
incorporate the formfactor in the Migdal resummation. Some of them
include formfactors also for the Migdal-vertices
\cite{Korpa1,Suzuki,Nakano}\footnote{Following the Korpa-Malfliet
  \cite{Korpa1,Korpa-priv} prescription for the Migdal vertices we
  could qualitatively reproduce their results as a check for our
  numerics.}.  We followed the approach from ref.  \cite{Urban} with
formfactor-free Migdal vertices.  The difference between these two
techniques turns out to have a great effect on the shape of the pion
spectral function for momenta in the order of 400 MeV.  For higher
temperatures all structures are smoothed out and the value of the
selfenergy even decreases because of limited number of available
nucleons to scatter.
%%%%%%%%%%%%%%%%%%%%%%%%%%%%%%%
\subsection{Pions and Vector-mesons}\label{Vec-mesons}
In this chapter we study the influence of the medium-modified
pion-modes on the spectral functions of the $\rho$- and
$\omega$-mesons.  The widths of these resonances are generated through
the decay into two pions in the case of the $\rho$-meson, respectively
three pions for the $\omega$-meson. For the $\omega$-meson we choose
the indirect decay via the $\rho$-meson (the so called Gell-Mann,
Sharp and Wagner (GSW) process \cite{GSW}) because, as shown by
Theileis \cite{Theileis}, this is the dominant contribution in vacuum.
The interaction Lagrangian defining the vertices is given by
\cite{Schwinger,Wess,Witten}
\begin{equation}\label{L-vector}
\Lag^{\rm{int}}_{\pi\rho\omega}
 =g_{\rho\pi\pi}\rho^\mu(\pi\overset{\leftrightarrow}{\partial_\mu}\pi)
  +g_{\omega\rho\pi}\epsilon^{\alpha\beta\mu\nu}
  \omega_\alpha\partial_\beta\rho_\mu\partial_\nu\pi,
\end{equation}
for a review see \cite{Klingl2}.  In this sector we are lead to the
following $\Phi$-functional approximation
\begin{fmffile}{PHI-meson}
\begin{equation}\label{PHI-meson}
\Phi=
\parbox{40\unitlength}{
\begin{fmfgraph*}(40,25)
        \fmfpen{thick}
        \fmfleft{i}
        \fmfright{o}
        \fmf{phantom}{i,v1}
        \fmf{phantom}{v2,o}
        \fmf{photon,left=1,tension=.15,label=$\rho$}{v1,v2}
        \fmf{dashes,right=1,tension=.15,label=$\pi$}{v1,v2}
        \fmf{gluon,tension=.15,label=$\omega$,l.d=10}{v1,v2}
\end{fmfgraph*}}
+
\parbox{40\unitlength}{
\begin{fmfgraph*}(40,25)
        \fmfpen{thick}
        \fmfleft{i}
        \fmfright{o}
        \fmf{phantom}{i,v1}
        \fmf{phantom}{v2,o}
        \fmf{dashes,left=1,tension=.15,label=$\pi$}{v1,v2}
        \fmf{dashes,right=1,tension=.15,label=$\pi$}{v1,v2}
        \fmf{photon,tension=.15,label=$\rho$,l.d=10}{v1,v2}
\end{fmfgraph*}}
\end{equation}
\end{fmffile}%.
From the point of view of a $\Phi$-derivable approximation we
neglected the coupling of the $\rho$-$\omega$-loop back to the pion.
This process is expected to contribute only marginally to the low
energy region of the pion spectrum, which turns out to be the most
relevant part to determine the width of the $\omega$-meson. Therefore
only the following retarded selfenergies are included:\\
$\phantom{aaaaaaaaa}$
\begin{fmffile}{SRho2Pi}
\begin{equation}\label{SRho2Pi}
\Pi_{\mu\nu\;\rho\pi\pi}^{R}=
\parbox{40\unitlength}{
\begin{fmfgraph*}(40,25)
        \fmfpen{thick}
        \fmfleft{i}
        \fmfright{o}
        \fmf{photon}{i,v1}
        \fmf{photon}{v2,o}
        \fmf{dashes,left=1,tension=.4,label=$\pi$}{v1,v2}
        \fmf{dashes,left=1,tension=.4,label=$\pi$}{v2,v1}
\end{fmfgraph*}}
\end{equation}
\end{fmffile}%
\begin{fmffile}{SOmegaRhoPi}
\begin{equation}\label{SOmegaRhoPi}
\Pi_{\mu\nu\;\omega\rho\pi}^{R}=
\parbox{40\unitlength}{
\begin{fmfgraph*}(40,25)
        \fmfpen{thick}
        \fmfleft{i}
        \fmfright{o}
        \fmf{gluon}{i,v1}
        \fmf{gluon}{v2,o}
        \fmf{dashes,left=1,tension=.4,label=$\pi$}{v2,v1}
        \fmf{photon,left=1,tension=.4,label=$\rho$}{v1,v2}
\end{fmfgraph*}}
\end{equation}
\end{fmffile}%
 The correlation between $\rho$- and
$\omega$-meson modes results from the $\omega$-meson selfenergy
(\ref{SOmegaRhoPi}) and the reverse process encoded in the coupling of
the $\rho$-meson to the $\omega$-$\pi$-loop
\begin{fmffile}{SRhoOmegaPi}
\begin{equation}\label{SRhoOmegaPi}
\Pi_{\mu\nu\;\rho\omega\pi}^{R}=
\parbox{40\unitlength}{
\begin{fmfgraph*}(40,25)
        \fmfpen{thick}
        \fmfleft{i}
        \fmfright{o}
        \fmf{photon}{i,v1}
        \fmf{photon}{v2,o}
        \fmf{dashes,left=1,tension=.4,label=$\pi$}{v2,v1}
        \fmf{gluon,left=1,tension=.4,label=$\omega$}{v1,v2}
\end{fmfgraph*}}
\end{equation}
\end{fmffile}%
which both are included selfconsistently.  Since the model omits higher lying
degrees of freedom the in-medium changes of the real parts of the
vector-meson selfenergies are less precisely determined within the
model space. Therefore we drop them, while restoring the normalisation
of the spectral functions at each iteration step in the further
analysis. In a later section we discuss the sensitivity of the results
on a possible change of the vector-meson masses in matter.

\subsection{Transversality of the vector-meson polarisation tensors}

The resulting polarisation-tensors (\ref{SRho2Pi} - \ref{SRhoOmegaPi})
have to be four-transversal because of current conservation.  In
normal perturbation theory this is guaranteed order by order.  In the
Dyson approach, where one sums up a restricted subclass of diagrams to
infinite order, one generally violates Ward identities on the
correlator level. Thus the polarisation tensor may contain
four-longitudinal components $\Pi_l^{\mu\nu}(q)$, which have to be
absent and which may lead to the propagation of unphysical degrees of
freedom. From general grounds, this deficiency can be cured by
corresponding vertex corrections. Without further approximations,
however, this leads to a presently intractable scheme of
Bethe-Salpeter equations which accounts for the required $t$-channel
exchanges required by crossing symmetry. We circumvent this problem in
the following way.  From transport considerations \cite{Knoll1} it is
known that such polarisation tensors have at least two relaxation
times. Because of charge conservation, one of these times has to be
infinite, implying that the component $\Pi^{00}(q)$ vanishes exactly
for $\textbf{q}=0$ and $q_0\neq 0$, while the second relaxation time
is clearly finite.  Such a result can never be reached in a truncated
Dyson-resummation scheme where all relaxation times are finite,
because they are determined by the damping-time scale of the dressed
propagators involved in the loops. On the other hand, the spatial
components of the polarisation tensor, given by the autocorrelation of
spatial currents, have solely finite and short correlation times which
can be expected to be safely approximated within a Dyson resummation
scheme.  Therefore our strategy assumes the spatial components of the
polarisation tensors $\Pi_{\mu\nu}^R$ to be given by the
selfconsistent loops, while the time-components are to be corrected
such that the full tensor becomes four-transversal. This is achieved
by using a projection technique where the full tensors (\ref{SRho2Pi}
- \ref{SRhoOmegaPi}) are decomposed into a four-longitudinal part
$\Pi_l$ and two four-transversal parts $\Pi_L$ and $\Pi_T$ which are
three-longitudinal and three-transversal, respectively,
\begin{equation}
\begin{split}
\label{projektion}
        \Pi^{\mu\nu}(q)&=\Pi_l^{\mu\nu}(q)+\Pi_L^{\mu\nu}(q)
        +\Pi_T^{\mu\nu}(q)\\   
        \Pi_L^{\mu\nu}(q)&
        =\left(-g^{\mu\nu}-\delta^{\mu\nu}+\frac{q^\mu q^\nu}{q^2}
          +\frac{\vec{q}^\mu \vec{q}^\nu}{\vec{q}^2}\right)\cdot\Pi_L(q)\\
        \Pi_T^{\mu\nu}(q)&=\left(\delta^{\mu\nu}
          -\frac{\vec{q}^\mu \vec{q}^\nu}{\vec{q}^2}\right)\cdot\Pi_T(q) .
\end{split}
\end{equation}
Here $g^{\mu\nu}$ is the metric tensor, whereas $\delta^{\mu\nu}$ and
$\vec{q}^{\mu}$ have vanishing time components.  We do not specify the
$\Pi_l$ part any further, since this part has just to be dropped
due to current conservation. The scalar functions $\Pi_L$
and $\Pi_T$ can be calculated solely from the spatial parts of the
polarisation tensors using the following traces
\begin{equation}
\begin{split}
        \Pi_1&=\frac{q^iq^k}{\vec{q}^2}\Pi^{ik}
        =\frac{(q^0)^2}{q^2}\cdot\Pi_L\\
        3\Pi_3&=\text{Tr}_3\left[\Pi^{ik}\right]
        =-g_{ik}\Pi^{ik}=2\Pi_T+\frac{(q^0)^2}{q^2}\cdot\Pi_L \quad\mbox{or}
        \label{Spur}\\
        \Pi_L&=\frac{q^2}{(q^0)^2}\cdot\Pi_1;\qquad\quad
        \Pi_T=\frac{1}{2}\left(3\Pi_3-\Pi_1\right).
\end{split}
\end{equation}
Due to the traces, the complicated Lorentz-structure of the interaction
vertices (\ref{L-vector}) in the loops can be reduced entirely to
expressions involving the four-, respectively, three-vector squares
$q^2$, $p^2$, ${p'}^2$ and $\vec{q}^2$, etc. of the external and
internal momenta involved in the loops. Furthermore, three-longitudinal
and transverse modes decouple such that the corresponding spectral
functions $A_{L}(q)$ and $A_{T}(q)$ are directly given by the scalar
functions $\Pi_{L}(q)$ and $\Pi_{T}(q)$ of the corresponding
components of the polarisation tensors.

\subsection{Vector-meson spectral functions in matter}

\begin{figure}[H]
        \begin{minipage}[t]{.48\linewidth}
         \epsfig{file=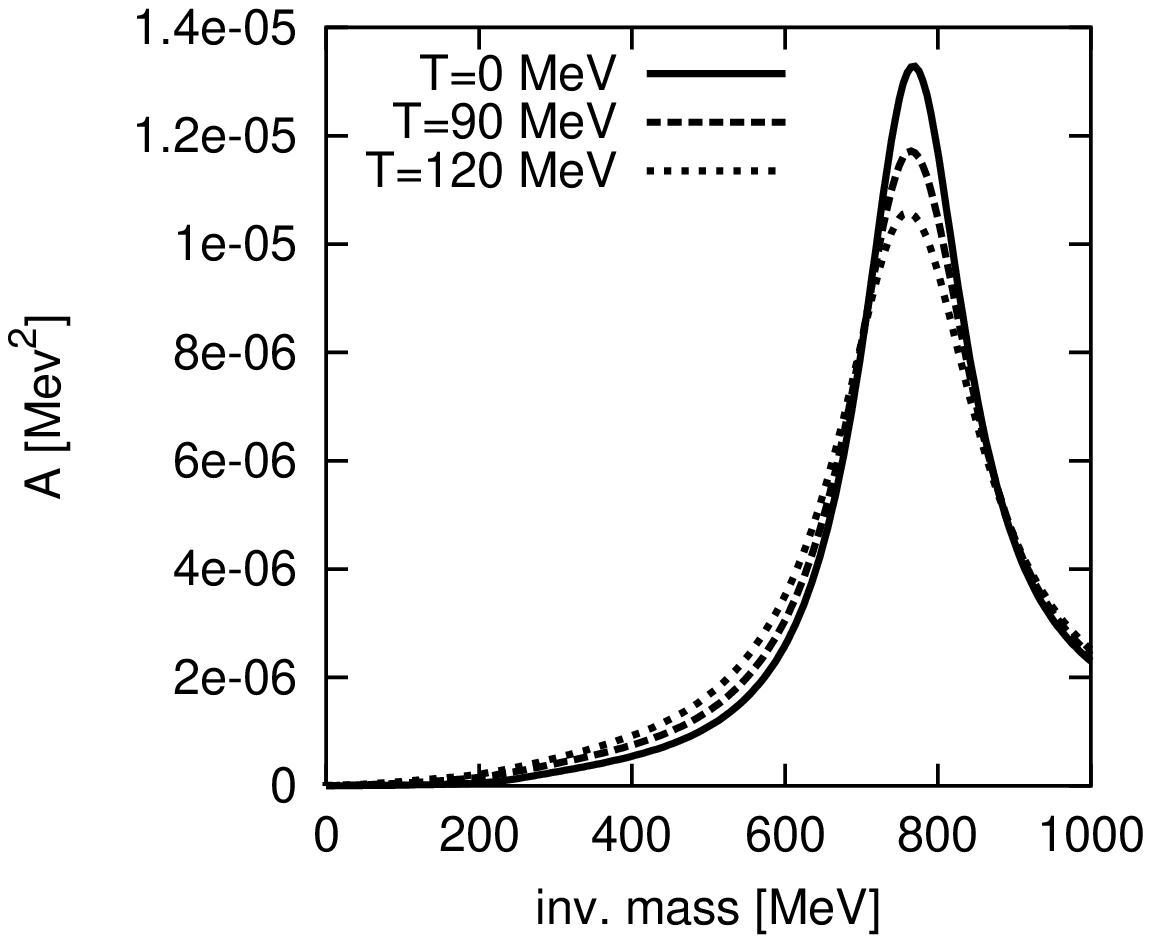,height=6.6 cm,width=6.0 cm,angle=-90}
         \caption{3-longitudinal $\rho$-meson spectral function
           $A_{\rho ,L}$ at p=200 MeV and $\rho =\rho_0$ for different
           temperatures\label{fig.5}} 
         \end{minipage}\hfill
        \begin{minipage}[t]{.48\linewidth}
         \epsfig{file=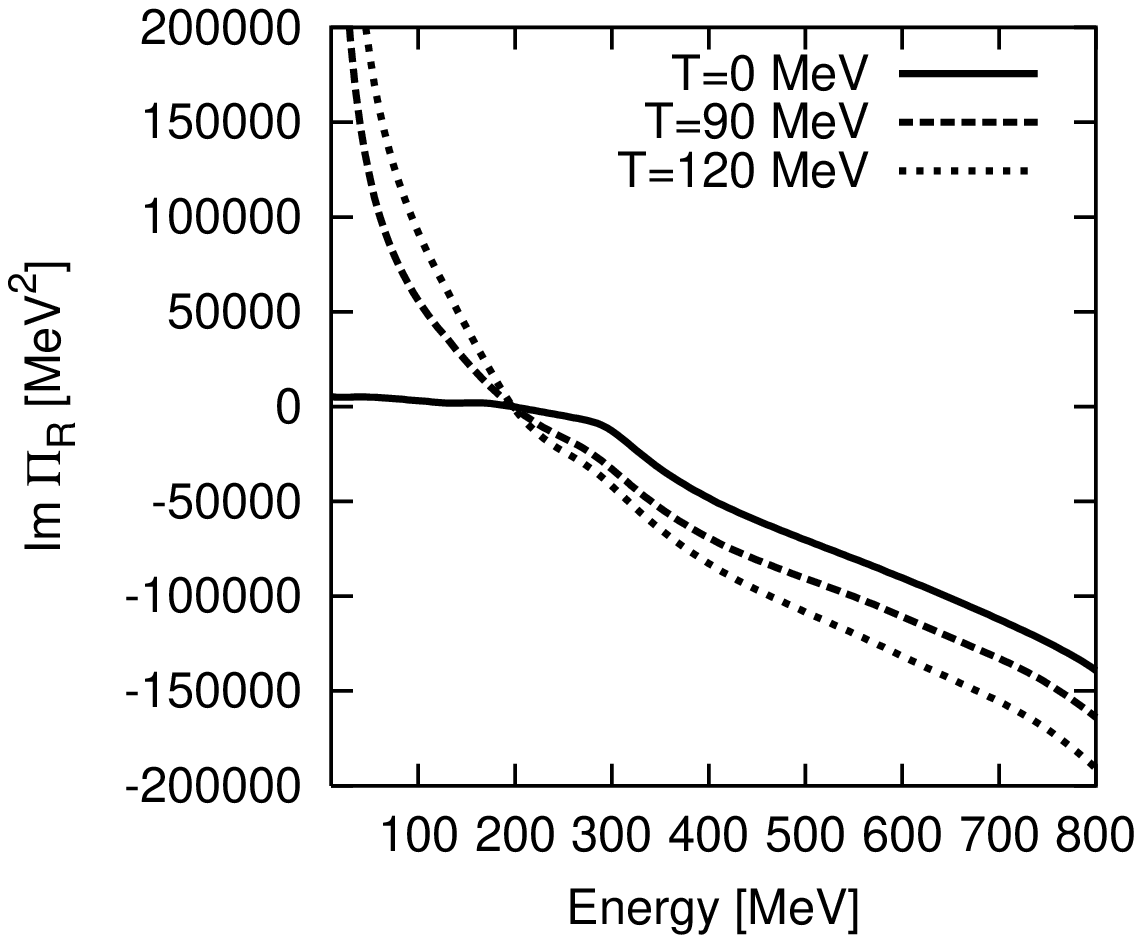,height=6.6 cm,width=6.0 cm,angle=-90}
         \caption{3-longitudinal part of the $\rho$-meson selfenergy $
           \Im\Pi^R_{\rho,L}$ at p=200 MeV and $\rho =\rho_0$ for
           different temperatures\label{fig.6}}         \end{minipage}
         \end{figure}

\begin{figure}[H]
        \begin{minipage}[t]{.48\linewidth}
         \epsfig{file=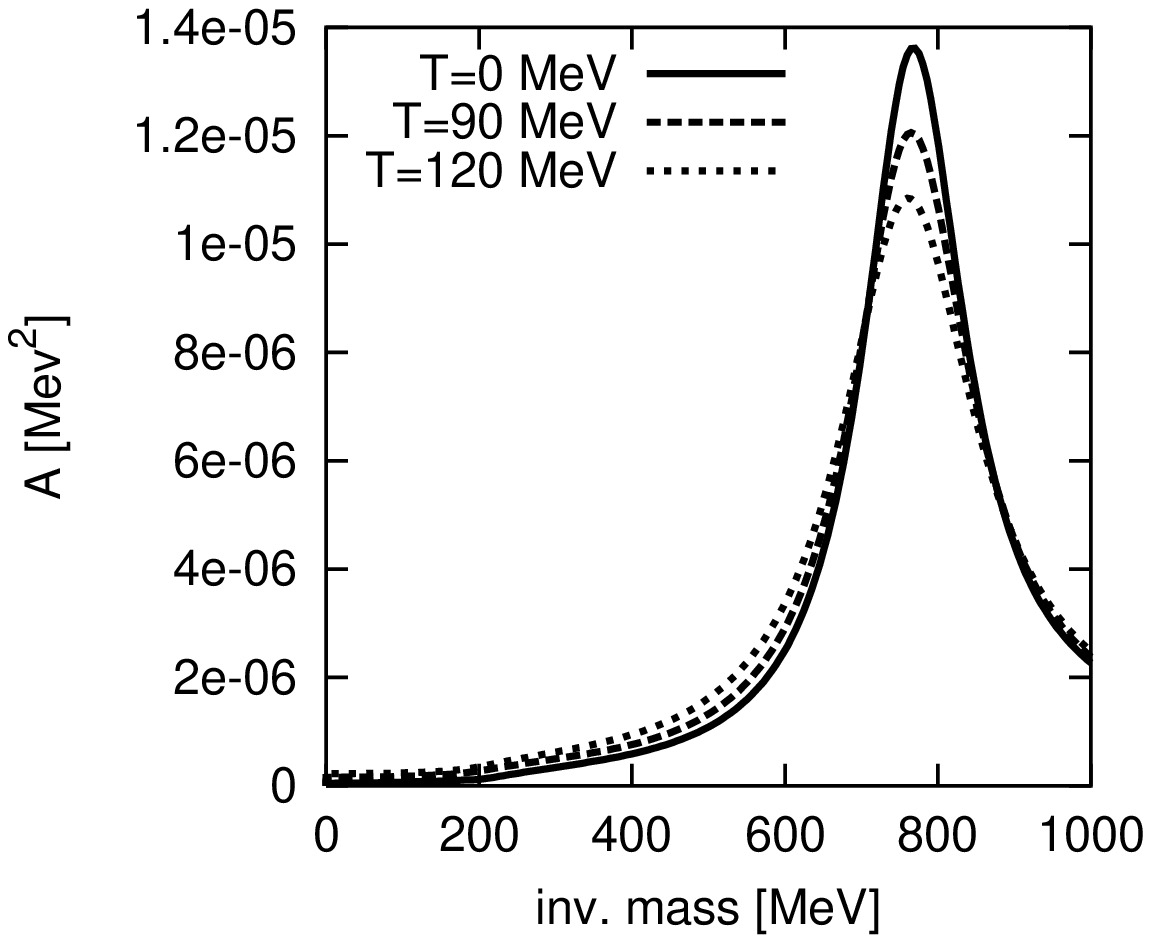,height=6.6 cm,width=6.0 cm,angle=-90}
         \caption{3-transversal $\rho$-meson spectral function $ A_{\rho ,T}$ 
          at p=200 MeV and $\rho =\rho_0$ for different temperatures}
         \end{minipage}\hfill
        \begin{minipage}[t]{.48\linewidth}
         \epsfig{file=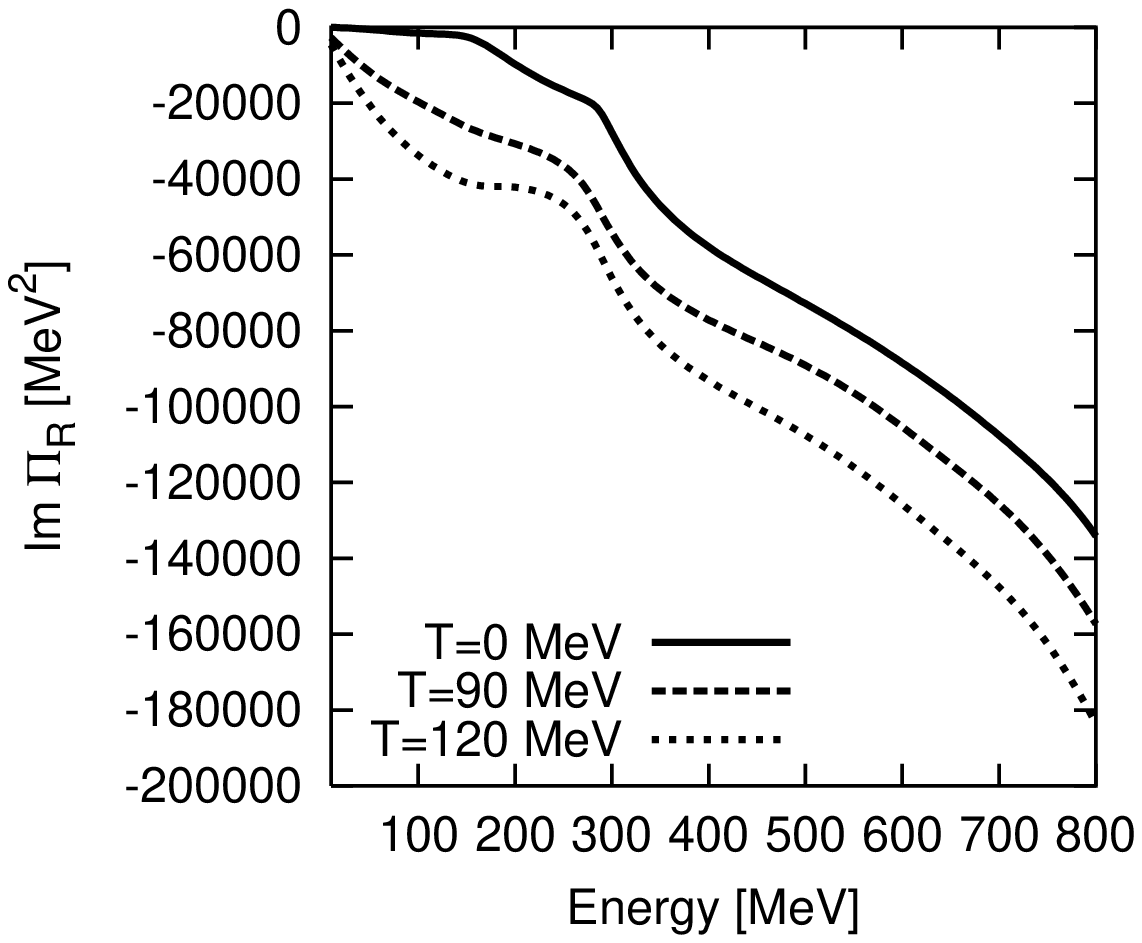,height=6.6 cm,width=6.0 cm,angle=-90}
         \caption{3-transversal part of the $\rho$-meson selfenergy $
           \Im\Pi^R_{\rho,T}$ at p=200 MeV and $\rho =\rho_0$ for
           different temperatures\label{fig.8} }
         \end{minipage}
         \end{figure}

\begin{figure}[H]
        \begin{minipage}[t]{.48\linewidth}
         \epsfig{file=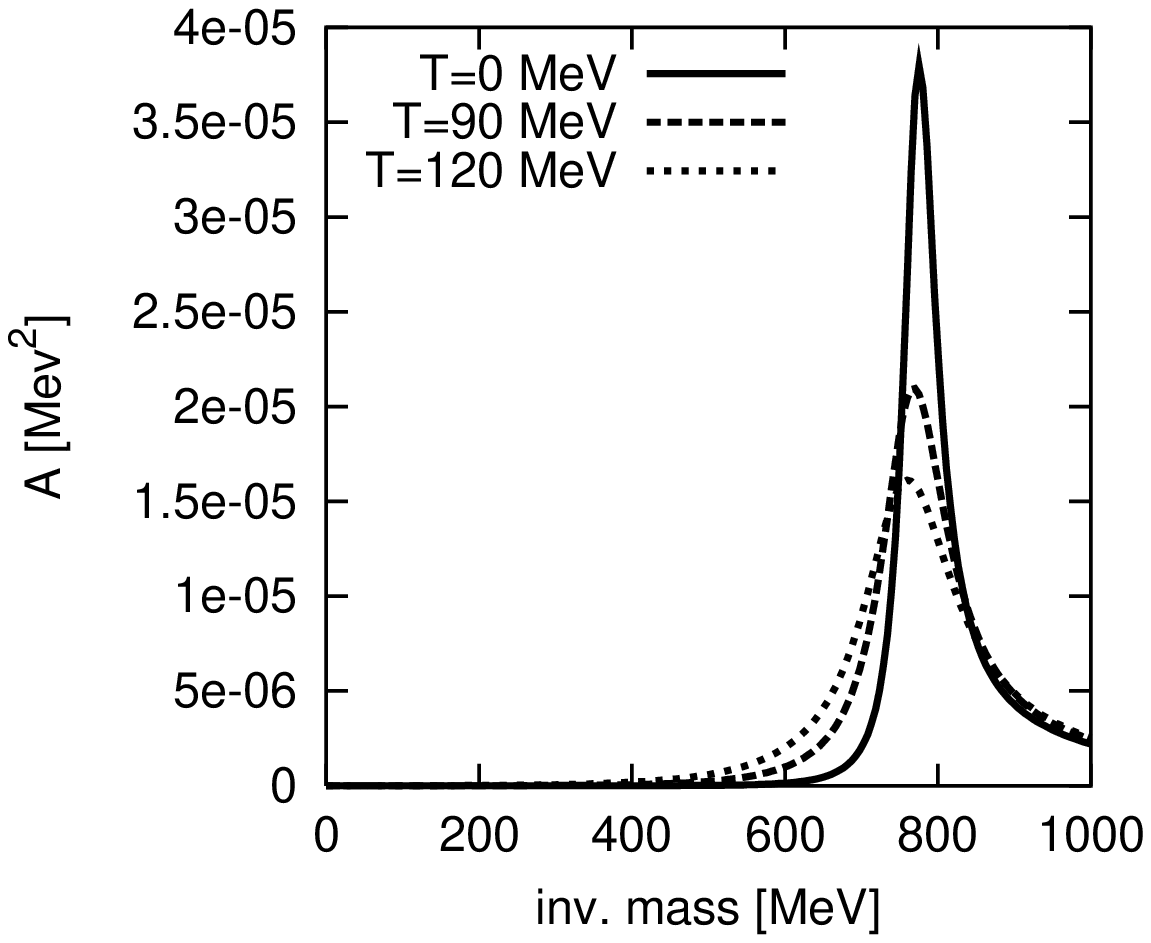,height=6.6 cm,width=6.0 cm,angle=-90}
         \caption{$A_{\omega ,L}$ at p=200 and $\rho =\rho_0$}
         \end{minipage}\hfill
        \begin{minipage}[t]{.48\linewidth}
         \epsfig{file=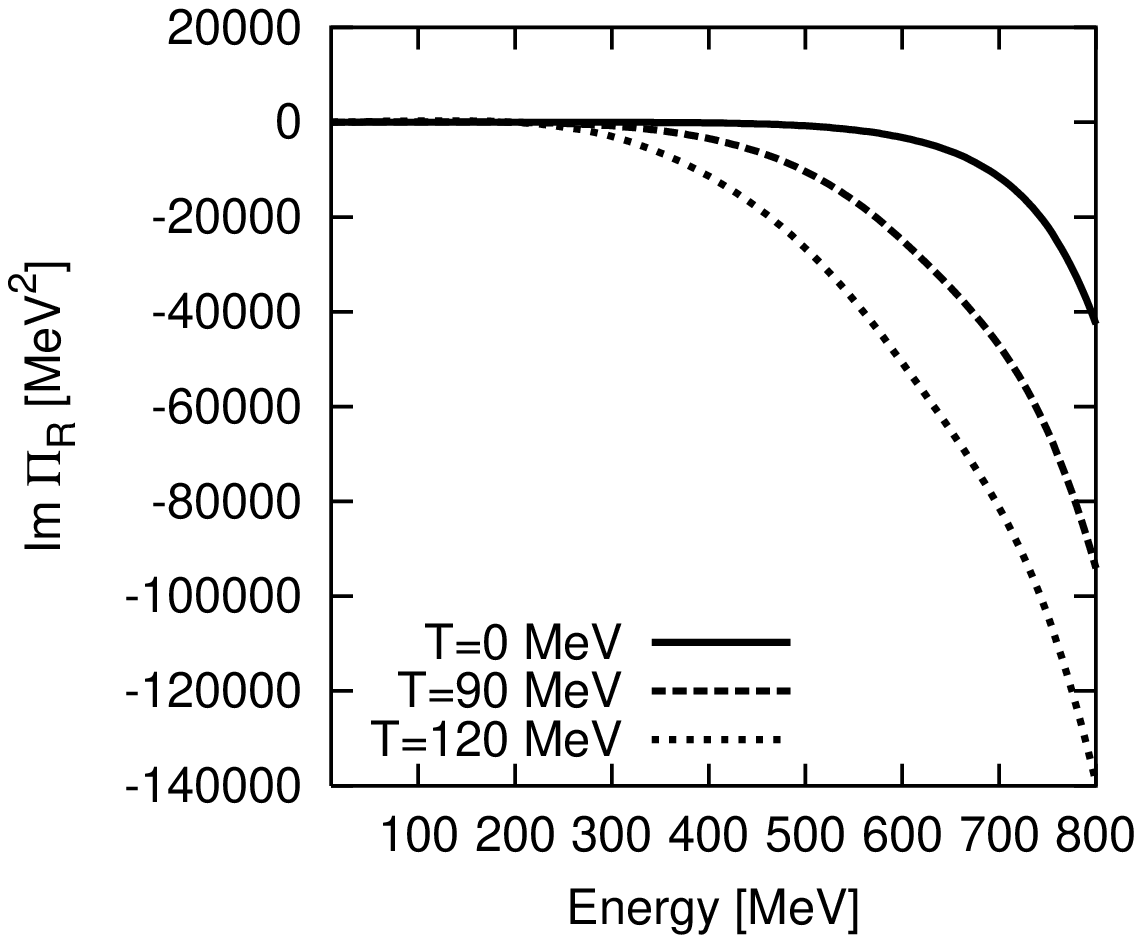,height=6.6 cm,width=6.0 cm,angle=-90}
         \caption{$\Im\Pi^R_{\omega,L}$ at p=200 MeV and $\rho
           =\rho_0$\label{fig.10}} 
         \end{minipage}
         \end{figure}

\begin{figure}[H]
        \begin{minipage}[t]{.48\linewidth}
         \epsfig{file=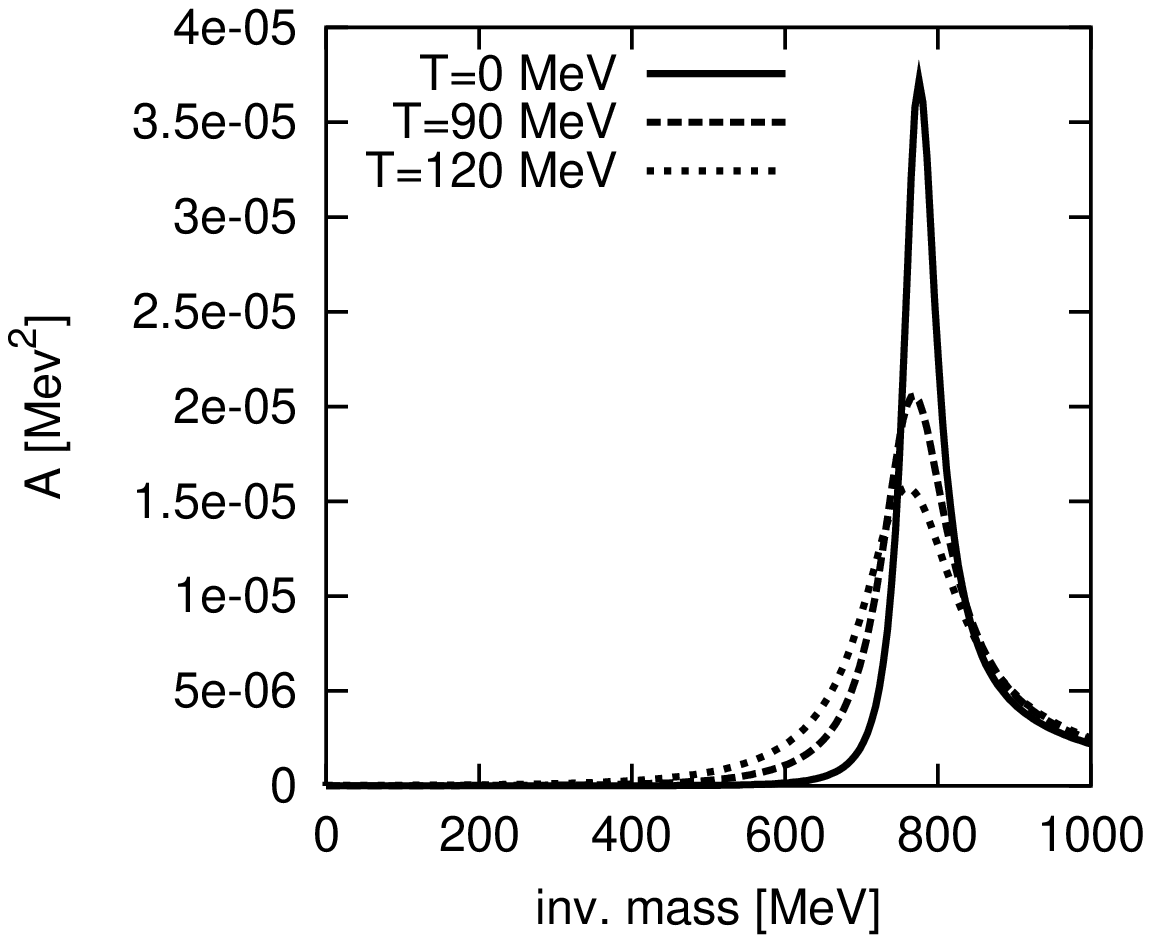,height=6.6 cm,width=6.0 cm,angle=-90}
         \caption{$A_{\omega ,T}$ at p=200 MeV and $\rho =\rho_0$}
         \end{minipage}\hfill
        \begin{minipage}[t]{.48\linewidth}
         \epsfig{file=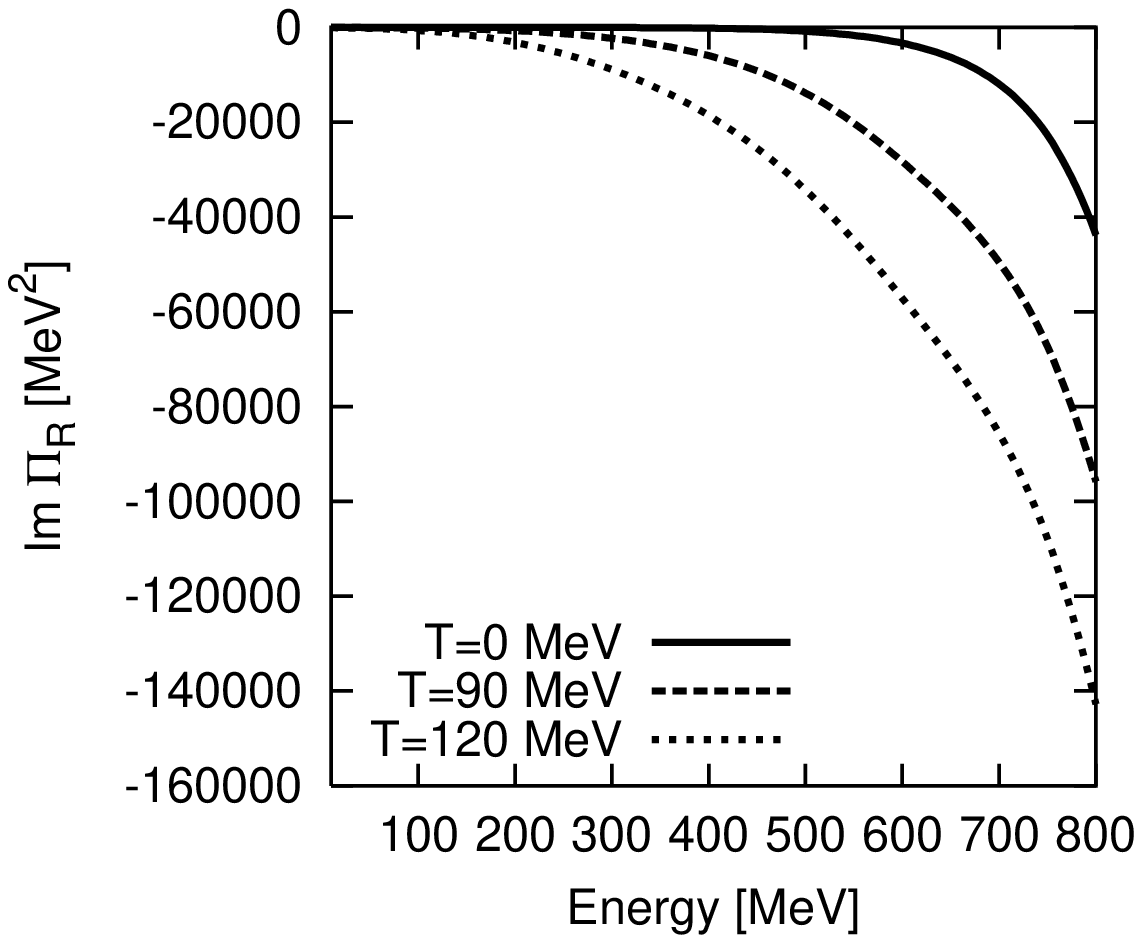,height=6.6 cm,width=6.0 cm,angle=-90}
         \caption{$\Im\Pi^R_{\omega,T}$ at p=200 MeV and $\rho
           =\rho_0$\label{fig.12}} 
         \end{minipage}
\end{figure}

With the dressed pion spectral function from section
\ref{subsect-piNDelta}, the three-longitudinal respectively
three-transversal polarisation tensors and spectral function of both
vector-mesons can now be calculated. In Figs. \ref{fig.5} to
\ref{fig.12} we show the results for normal nuclear density at
different temperatures\footnote{Please note that by definition, cf.
  (\ref{projektion}),  $\Im\Pi^R_{L}$ vanishes on the light cone
  $p^2=0$ and becomes positive in the space-like region.}


The main effect towards finite densities and/or finite temperatures is
the disappearance of thresholds present in the free particle
kinematics for the pions in the loops. The spectral strength starts
right at zero energy for both vector-mesons, though less visible for
the $\omega$-meson due to the smallness of the coupling.  This has
significant consequences for the low mass region of the corresponding
dilepton yields (cf. next section).  The temperature has no great
influence on the $\rho$-meson spectrum, while the $\omega$-meson width
increases significantly. For the latter effect indirectly the
nucleon--nucleon-hole excitations are of special importance as
discussed later.
         \begin{figure}[H]
        \begin{minipage}[t]{.48\linewidth}
         \epsfig{file=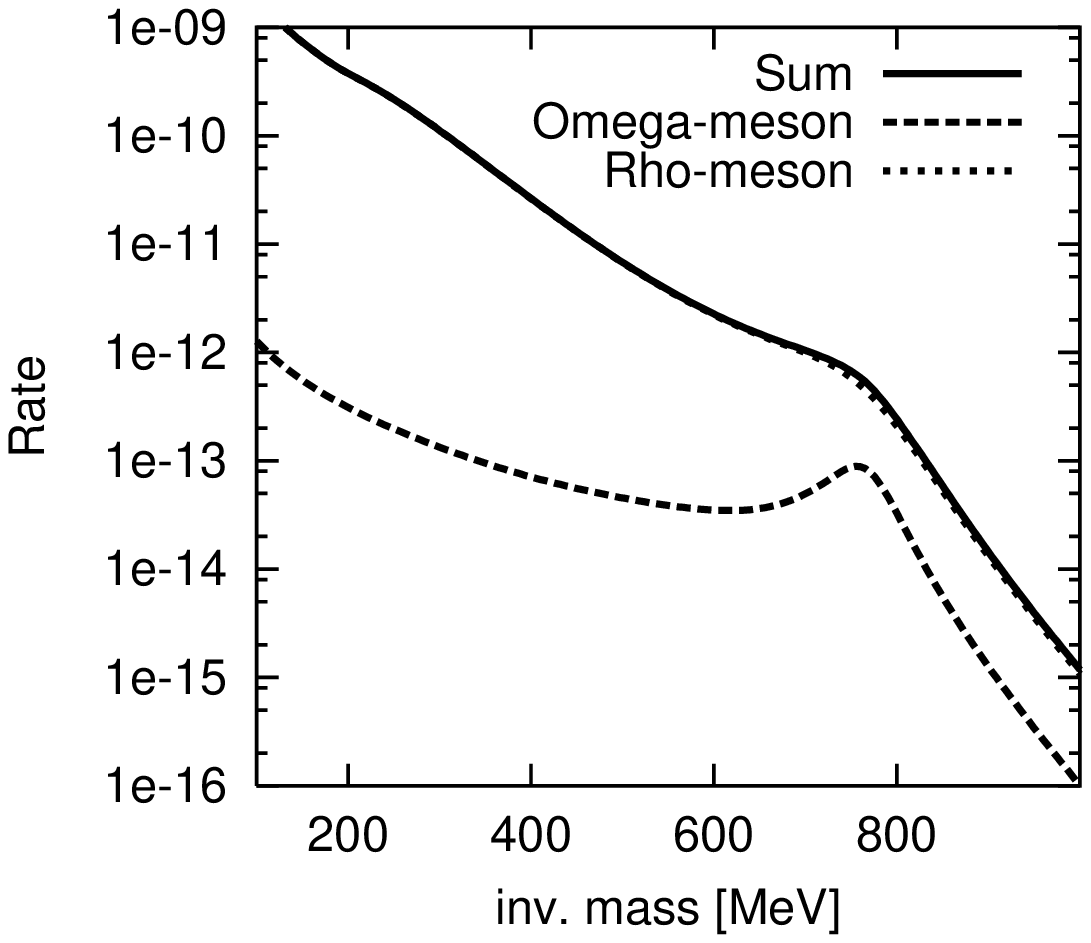,height=6.6 cm,width=6.0 cm,angle=-90}
         \caption{$e^+e^-$-rate at T=60 MeV $\rho=\rho_0$\label{fig13}}
         \end{minipage}\hfill
        \begin{minipage}[t]{.48\linewidth}
         \epsfig{file=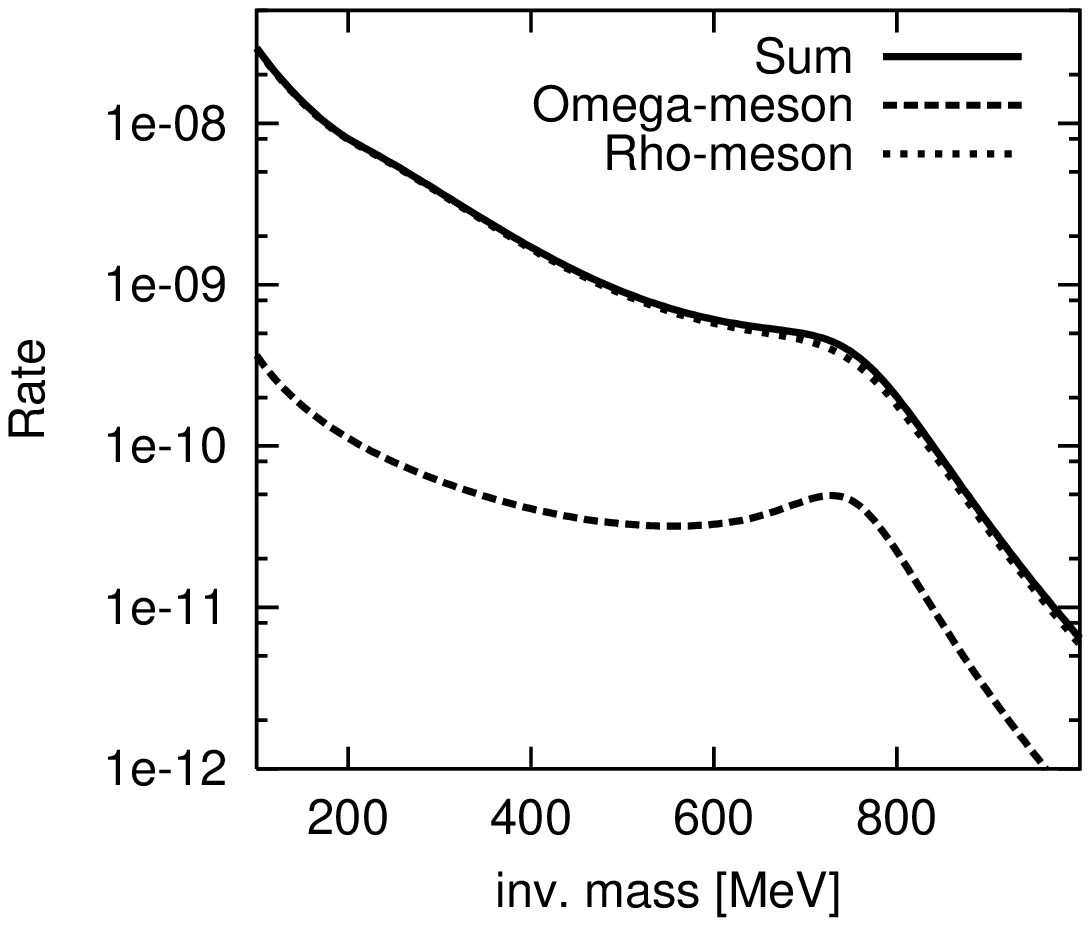,height=6.6 cm,width=6.0 cm,angle=-90}
         \caption{$e^+e^-$-rate at T=120 MeV $\rho=\rho_0$}
         \end{minipage}\\
        \begin{minipage}[t]{.48\linewidth}
         \epsfig{file=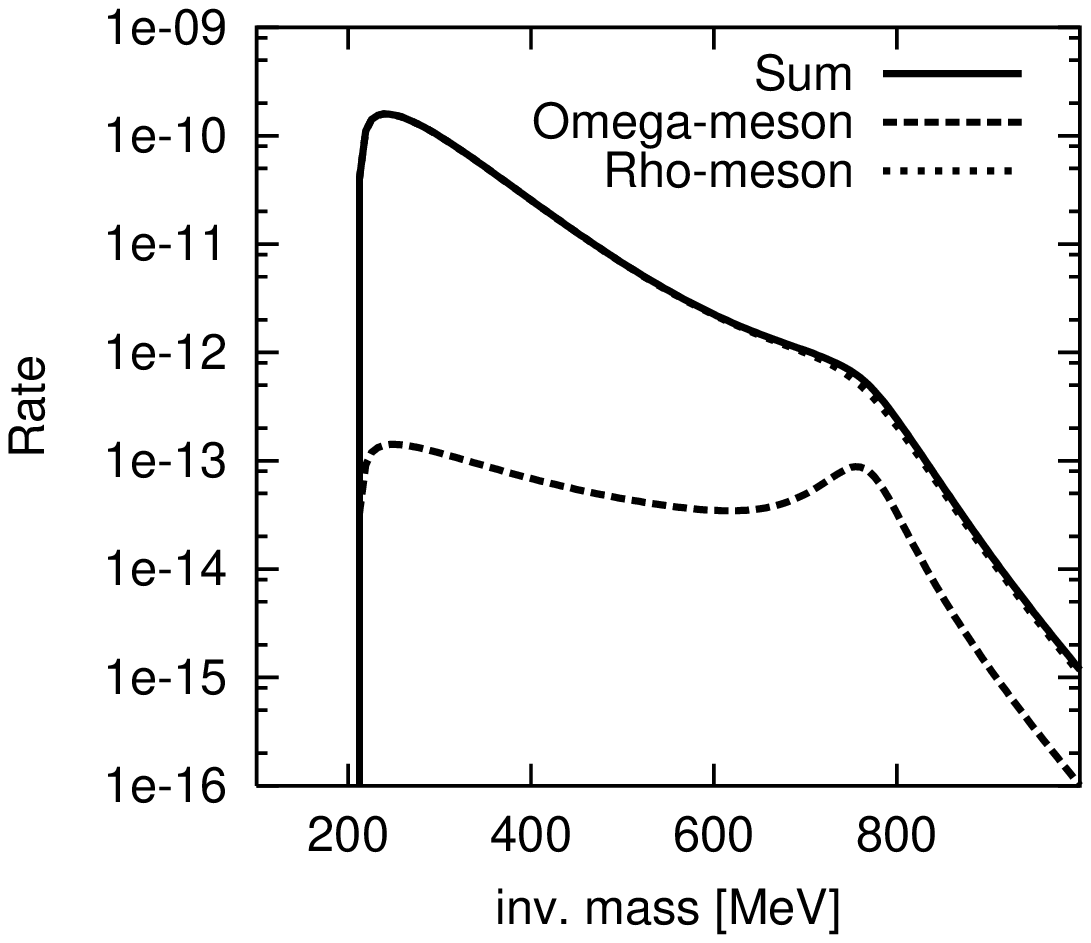,height=6.6 cm,width=6.0 cm,angle=-90}
         \caption{$\mu^+\mu^-$-rate at T=60 MeV $\rho=\rho_0$}
         \end{minipage}\hfill
        \begin{minipage}[t]{.48\linewidth}
         \epsfig{file=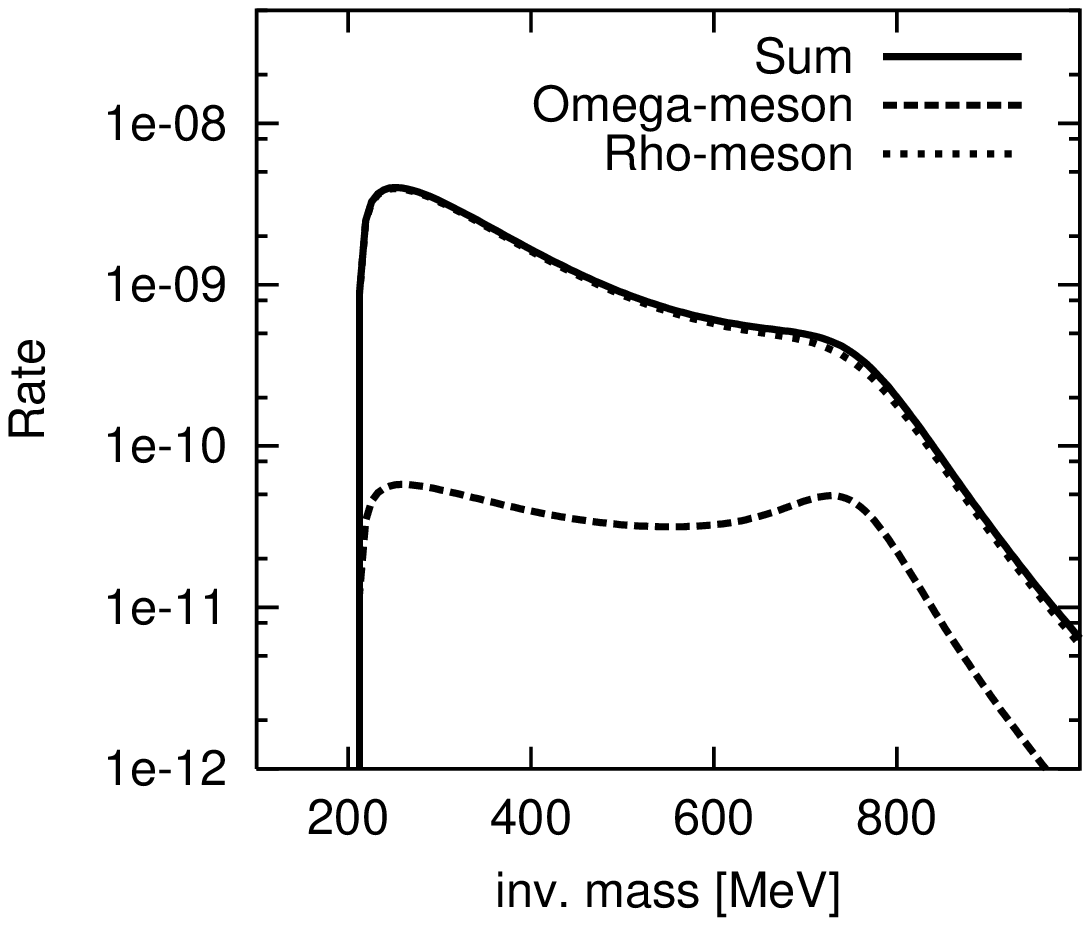,height=6.6 cm,width=6.0 cm,angle=-90}
         \caption{$\mu^+\mu^-$-rate at T=120 MeV $\rho=\rho_0$\label{fig16}}
         \end{minipage}
\end{figure}

\subsection{Dileptons}\label{Dileptons}
Using vector-dominance we can calculate \cite{Sakurai,Klingl}
the production-rate for dilepton pairs ($e^+e^-$ and $\mu^+\mu^-$)
\begin{equation}
\begin{split}
\frac{dR}{d^4qd^3xdt}=
&\frac{\alpha^2}{6\pi^3q^2}
\sqrt{\frac{q^2-4m_l^2}{q^2}}(1+\frac{2m_l^2}{q^2})\\
&\times\sum_{\rm{v}}\frac{m_{\rm{v}}^4}{g^2_{\rm{v}\pi\pi}(1-4\pi\alpha
  g^2_{\rm{v}\pi\pi})} 
[2A_{\rm{v},T}(q,T)+A_{\rm{v},L}(q,T)]n_B(q,T)
\end{split}
\end{equation}
from the in-medium spectral functions of the $\rho$- and
$\omega$-meson.  Here $n_B(q_0,T)$ is the thermal Bose-Einstein
weight, while  $m_{\rm{v}}$ and $g_{\rm{v}\pi\pi}$ denote the vector-meson mass 
 and the coupling constant of the meson to two pions. The
mass of the lepton, electron or muon, is symbolised by $m_l$. In our
model we obtain the results displayed in Figs. \ref{fig13} to
\ref{fig16} for the direct decay of a vector-resonance into a dilepton
pair via an intermediate time-like photon.

With applications to nuclear collisions in mind, we use the vacuum
expressions for the photon- and lepton-propagators, since these
particles interact only weakly, such that they do not get modified by
the surrounding medium. In addition, we used the coupling constants
from \cite{Klingl2} in order to describe the partial decay width of
both mesons into dileptons.  One can see that, due to the thermal
weight and the broad spectral functions, the dilepton spectra are
essentially falling over the whole energy range.  This would clearly
be different if the vacuum spectral function for the $\omega$-meson
had been used here. The $\mu^+\mu^-$-spectrum has a threshold at twice
the muon mass. At higher invariant mass both di-lepton spectra merge,
because then the muon mass becomes negligible in comparison with the
energy. The contributions of the $\omega$-meson are hardly visible in
the total spectra because of its increased width. For equal widths of
both mesons the contributions of the $\omega$-meson falls below that of
the $\rho$-meson by one order of magnitude because of the smaller
partial decay width into dileptons.

In the following we explicitly analyse the influence of the various
components of the pion spectral function on the vector-mesons and thus
on the dilepton spectra. Formally we do this by splitting the spectral
function of the vector-meson into the various components related to
the different processes feeding into this vector-meson channel. Thus,
decomposing the total damping width into partial widths
$\Gamma_{\rm{v},\rm{tot}}(p)=\sum_i\Gamma_{\rm{v},\rm{i}}(p)$ the
dilepton yield can be brought into a Breit-Wigner like form with
partial in- and out-widths
\begin{equation}
\begin{split}
\frac{d R}{d^4q d^3xdt}
       &=\frac{3}{(2\pi)^4}\;n_T(q_0)\sum_{\rm{v}}A_{\rm{v}}(q)\;(-2\Im
       \Pi_{\rm{v},\rm{e}^+\rm{e^-}})\cr  
       &=\frac{3}{(2\pi)^4}\;n_T(q_0)\sum_{\rm{v},i}
       \frac {4\;q_0^2\Gamma_{\rm{v},\rm{i}}\;\Gamma_{\rm{v},\rm{e}^+\rm{e}^-}}
             {(q^2-m_{\rm{v}}^2)^2+q_0^2\Gamma_{\rm{v},\rm{tot}}^2}
\end{split}\label{d R}
\end{equation}
(suppressing the tensor structure of spectral function and
polarisation tensor which leads to a degeneracy factor 3 for vector
particles).  Here, $\Gamma_{\rm{v},\rm{e}^+\rm{e}^-}$ is the dilepton
decay width of vector-meson $\rm{v}$.

We start the discussion of the effects on the $\omega$-meson. In the
medium there are three major processes contributing to its damping
width, illustrated by perturbative ``time-flow'' diagrams where the
time is running from left to right and vertical lines denote a virtual
space-like propagator, which mediates a two-body interaction:\\

\begin{figure}[t]
\begin{minipage}[t]{.48\linewidth}
\epsfig{file=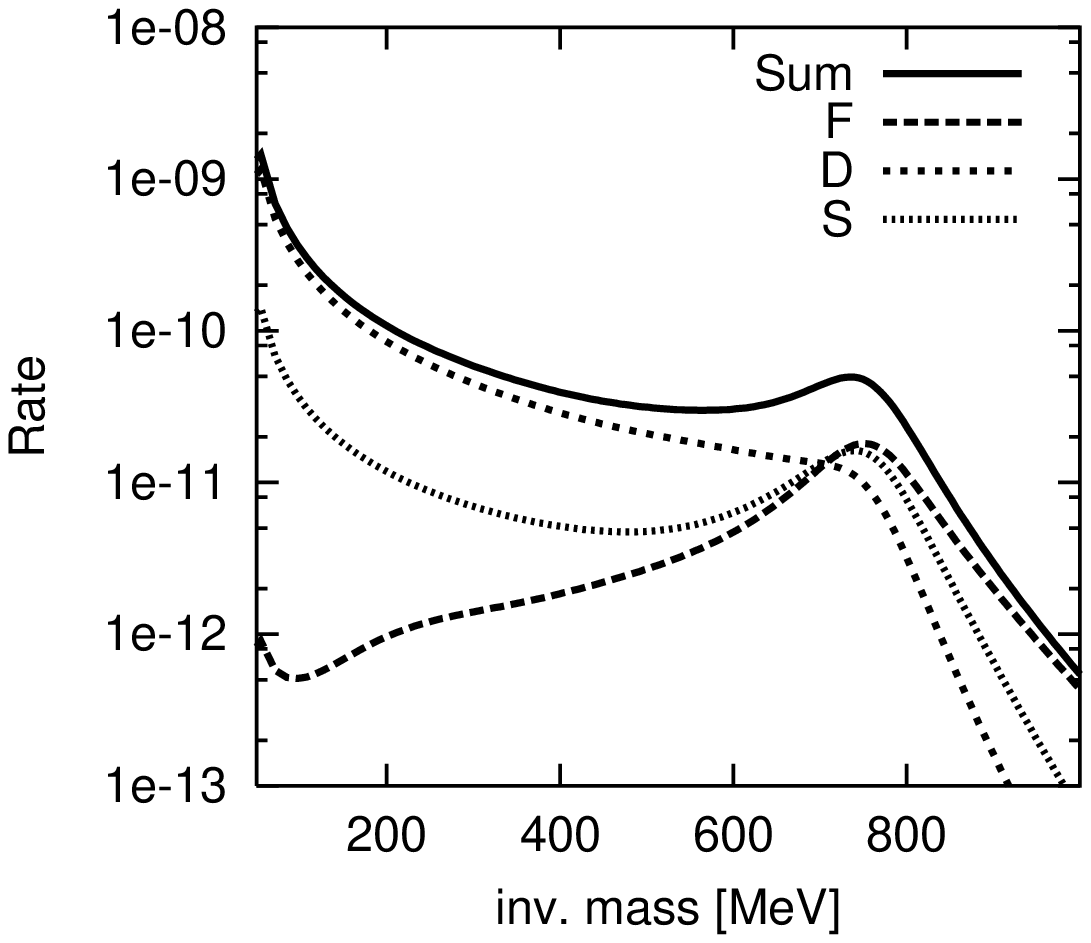,height=6.6 cm,width=6.0
  cm,angle=-90} 
\caption{$e^+e^-$-rate from the decay of the $\omega$-meson at T=120
  MeV $\rho=\rho_0$ divided into different contributions:
\label{omega-div}}
\begin{tabular}{llll}
F:&$\pi\rho\rightarrow\omega$\\[-3mm]
D:&$\rho\rightarrow\pi\omega$\\[-3mm]
S:&$\rho N\rightarrow\omega N$
\end{tabular}
\end{minipage}\hfill
\begin{minipage}[t]{.48\linewidth}
\epsfig{file=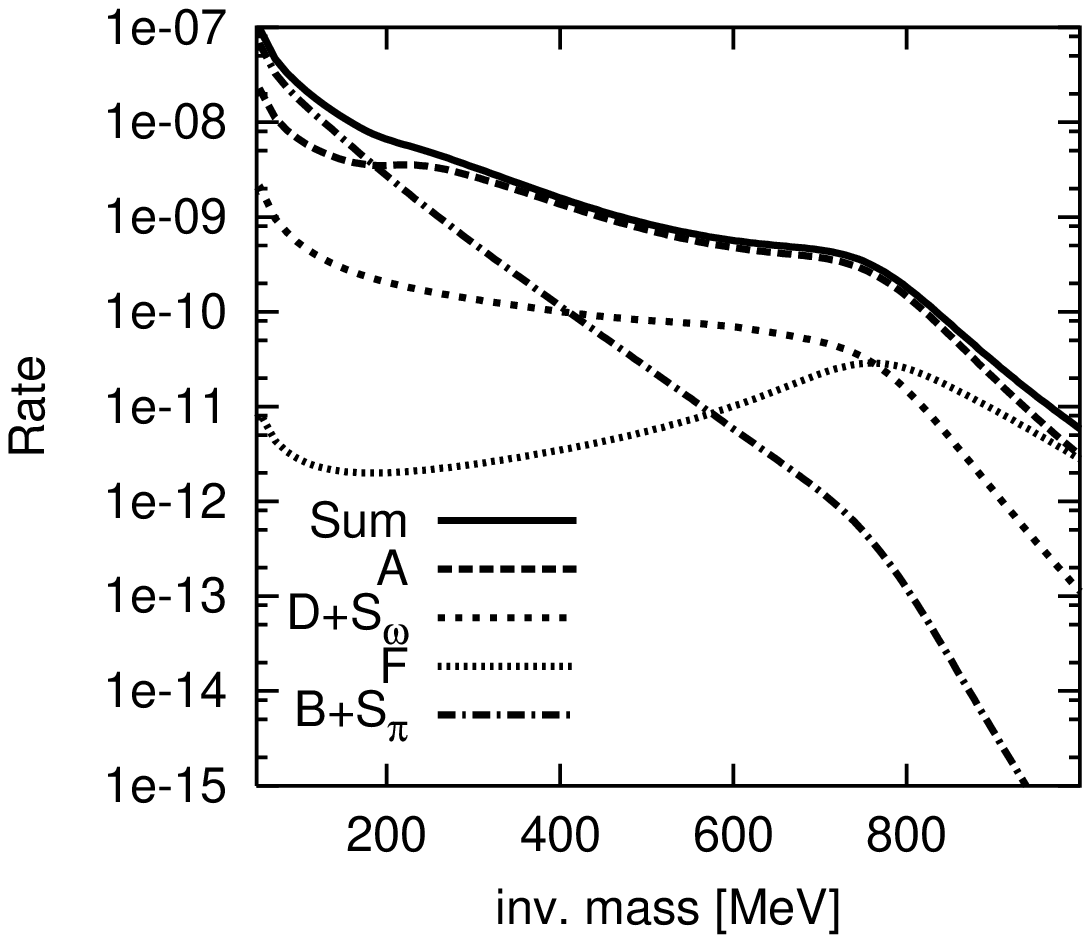,height=6.6 cm,width=6.0 cm,angle=-90}
\caption{$e^+e^-$-rate from the decay of the 
$\rho$-meson at T=120 MeV $\rho=\rho_0$ divided into  different contributions:
\label{rho-mix}}
\begin{tabular}{llll}
A:&$\pi\pi\rightarrow\rho$&F:&$\pi\omega\rightarrow\rho$\\[-3mm]
B:&$\pi\rightarrow\pi\rho$&D:&$\omega\rightarrow\pi\rho$\\[-3mm]
S$_{\pi}$:&$\pi N\rightarrow\rho N$&S$_{\omega}$:&$\omega
N\rightarrow\rho N$
\end{tabular}
\end{minipage}
\end{figure}
%%%%%%%%%%%%%%%%%%%%%%%%%%%%%%%%%%%%%%%%%%%%
\setlength{\unitlength}{0.5mm}
\begin{equation}
\begin{fmffile}{OmegaRhoScat}
(\rm{F})=\parbox{40\unitlength}{
\begin{fmfgraph*}(40,30)
        \fmfpen{thin}
        \fmfleft{i,i2}
        \fmfright{o}
        \fmf{dashes}{i,v1}
        \fmf{gluon}{v1,o}
        \fmf{photon}{i2,v1}
        \fmflabel{$\pi$}{i}
        \fmflabel{$\rho$}{i2}
        \fmflabel{$\omega$}{o}
\end{fmfgraph*}}
\end{fmffile}
\hspace*{1cm}
(\rm{D})=\quad\begin{fmffile}{OmegaRhoPair}
\parbox{40\unitlength}{
\begin{fmfgraph*}(40,30)
        \fmfpen{thin}
        \fmfleft{i}
        \fmfright{o,o2}
        \fmf{gluon}{v1,o2}
        \fmf{dashes}{v1,o}
        \fmf{photon}{i,v1}
        \fmflabel{$\rho$}{i}
        \fmflabel{$\pi$}{o}
        \fmflabel{$\omega$}{o2}
\end{fmfgraph*}}
\end{fmffile}
\hspace*{1cm}
\begin{fmffile}{OmegaPiScat}
(\rm{S})=\parbox{40\unitlength}{
\begin{fmfgraph*}(40,30)
        \fmfpen{thin}
        \fmfleft{i,i2}
        \fmfright{o,o2}
        \fmf{fermion}{i,v1}
        \fmf{fermion}{v1,o}
        \fmflabel{$N$}{i}
        \fmflabel{$N$}{o}
        \fmf{dashes,label=$\pi$,l.d=2}{v2,v1}
        \fmf{photon}{i2,v2}
        \fmf{gluon}{v2,o2}
        \fmflabel{$\rho$}{i2}
        \fmflabel{$\omega$}{o2}
\end{fmfgraph*}}
\end{fmffile}
\label{om-processes}
\end{equation}\\
\setlength{\unitlength}{1mm}%
Here the subsequent decay of the $\omega$-meson into the virtual
time-like photon and its final decay into the lepton pair is not
illustrated. In the selfconsistent calculation all these processes are
included automatically by using dressed propagators. For the first
process (F), $\rho\pi\rightarrow \omega\rightarrow \rm{e}^+\rm{e}^-$,
the $\omega$-meson is formed by the fusion of a $\rho$-meson with a
quasi-real, in-medium pion. Its inverse exists already in vacuum and
determines the vacuum decay width of the $\omega$-meson. The second
process (D), $\rho\rightarrow \pi\omega\rightarrow\pi
\rm{e}^+\rm{e}^-$, corresponds to a $\rho$-Dalitz-decay via an
intermediate $\omega$-meson.  In the selfconsistent calculations, both
above mentioned processes just differ in the sign of the pion energy
in the $\pi\rho$-loop of the $\omega$-selfenergy (\ref{SOmegaRhoPi}).
The process (S) in (\ref{om-processes}) corresponds to the
scattering $\rho$N$\rightarrow\omega$N mediated by a virtual, i.e.
space-like pion exchange. In view of the pion modes at zero
temperature (cf. Fig.  \ref{fig1}) we isolate this space-like
component by a cut on the far space-like region with pion loop momenta
with $|\vec{p}|>2|p^0|$ in (\ref{SOmegaRhoPi}).  At $T=120$ MeV this
separation is less evident in view of the broad structure of the pion
spectral function (Fig. \ref{pion-spec}). Thus the different
components of the processes displayed in Fig.  \ref{omega-div}
somewhat depend on this cut. In addition to the fusion width (F),
which constitutes just a temperature dependent modification of the
vacuum width (e.g. accounted for by Schneider and Weise
\cite{Schneider1}) a genuine in-medium process, namely the scattering
process (S), contributes with comparable strength at the nominal
resonance position. The ``Daliz''-decay of the $\rho$-meson (D)
dominates the low mass region. 

In summary of this analysis: a major portion to the $\omega$
spectrum results from processes (S) which are not accounted for in the
simple on-shell treatment. These contributions, however, sensitively
depend on the in-medium properties of the virtual pion cloud, which
certainly needs further clarifying investigations before quantitative
conclusions can be drawn.

%%%%%%%%%%%%%%%%%%%%%%%%%%%%%%%
The same type of processes as in (\ref{om-processes}) also occur for
the $\rho$-meson just interchanging $\rho$ with $\omega$, listed in
the $\rho$-meson decomposition given in Fig. \ref{rho-mix} as (F), (D)
and (S$_{\omega}$). However, these $\omega$-induced components are less
important compared to the coupling to the two-pion channels

\setlength{\unitlength}{0.5mm}
\begin{equation}
\begin{fmffile}{RhoPiPair}
(\rm{A})=\parbox{40\unitlength}{
\begin{fmfgraph*}(40,30)
        \fmfpen{thin}
        \fmfleft{i,i2}
        \fmfright{o}
        \fmf{dashes}{i,v1}
        \fmf{photon}{v1,o}
        \fmf{dashes}{i2,v1}
        \fmflabel{$\pi$}{i}
        \fmflabel{$\pi$}{i2}
        \fmflabel{$\rho$}{o}
\end{fmfgraph*}}
\end{fmffile}
\hspace*{1cm}
(\rm{B})=\quad\begin{fmffile}{RhoPiBrems}
\parbox{40\unitlength}{
\begin{fmfgraph*}(40,30)
        \fmfpen{thin}
        \fmfleft{i}
        \fmfright{o,o2}
        \fmf{photon}{v1,o2}
        \fmf{dashes}{v1,o}
        \fmf{dashes}{i,v1}
        \fmflabel{$\pi$}{i}
        \fmflabel{$\pi$}{o}
        \fmflabel{$\rho$}{o2}
\end{fmfgraph*}}
\end{fmffile}
\hspace*{1cm}
\begin{fmffile}{PiNRhoNScat}
(\rm{S}_{\pi})=\parbox{40\unitlength}{
\begin{fmfgraph*}(40,30)
        \fmfpen{thin}
        \fmfleft{i,i2}
        \fmfright{o,o2}
        \fmf{fermion}{i,v1}
        \fmf{fermion}{v1,o}
        \fmflabel{$N$}{i}
        \fmflabel{$N$}{o}
        \fmf{dashes,label=$\pi$,l.d=2}{v2,v1}
        \fmf{dashes}{i2,v2}
        \fmf{photon}{v2,o2}
        \fmflabel{$\pi$}{i2}
        \fmflabel{$\rho$}{o2}
\end{fmfgraph*}}
\end{fmffile}
\label{rho-processes}
\end{equation}\\
\setlength{\unitlength}{1mm}%
 At invariant masses above 300 MeV, the
$\pi^+\pi^-$-annihilation process (A) is clearly dominant. Processes
(B) and ($S_\pi$) are not present in any on-shell treatment of the pion as
they solely arise from genuine off-shell components of the pion
spectral function.  In this respect (B) can be interpreted as a
bremsstrahlung process radiated off a pion scattered in the medium,
while process (S) corresponds to inelastic $\pi N\rightarrow \rho N$
scatterings mediated via virtual pion exchange. The latter two
components, which emerge completely consistently within the model,
only contribute to the very low-mass part of the invariant mass
spectrum.
%%%%%%%%%%%%%%%%%%%%%%%%%%%%%%%
%%%%%%%%%%%%%%%%%%%%%%%%%%%%%%%%%%%%%%%%%%%%%%%%%%%%%%%%%%%%

\subsection{Dependence on density and temperature}

\begin{figure}[t]
        \begin{minipage}[t]{.48\linewidth}
         \epsfig{file=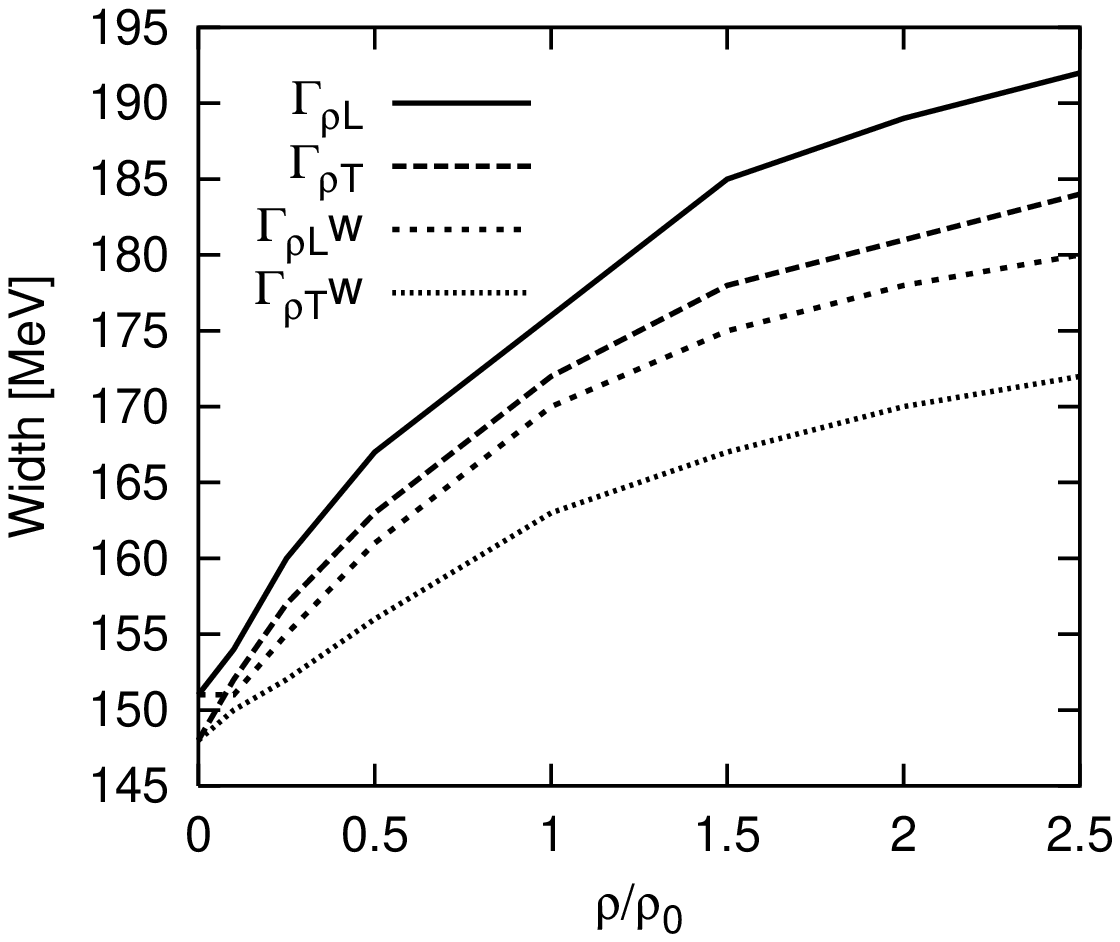,height=6.6
         cm,width=5.6 cm,angle=-90}
         \caption{$\Gamma_\rho$ at T=30 MeV depending on the
         density. $\Gamma$ w marks the process with only
         $\rho\rightarrow\pi\pi$.\label{Breite-Rho-Dichte-mit}} 
         \end{minipage}\hfill
        \begin{minipage}[t]{.48\linewidth}
          \epsfig{file=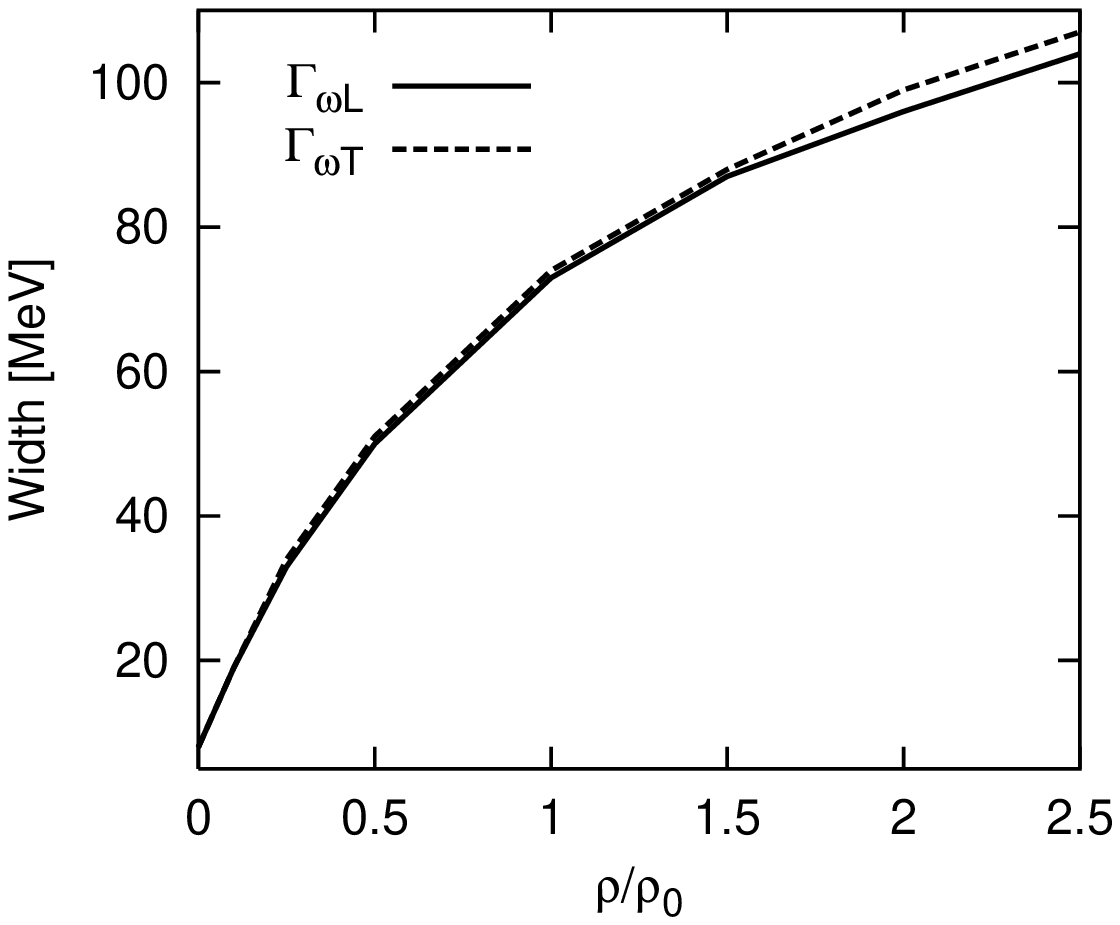,height=6.6
            cm,width=5.6 cm,angle=-90}
         \caption{$\Gamma_\omega$ at T=30 MeV depending on the density.}
         \end{minipage}
         \\
        \begin{minipage}[t]{.48\linewidth}
         \epsfig{file=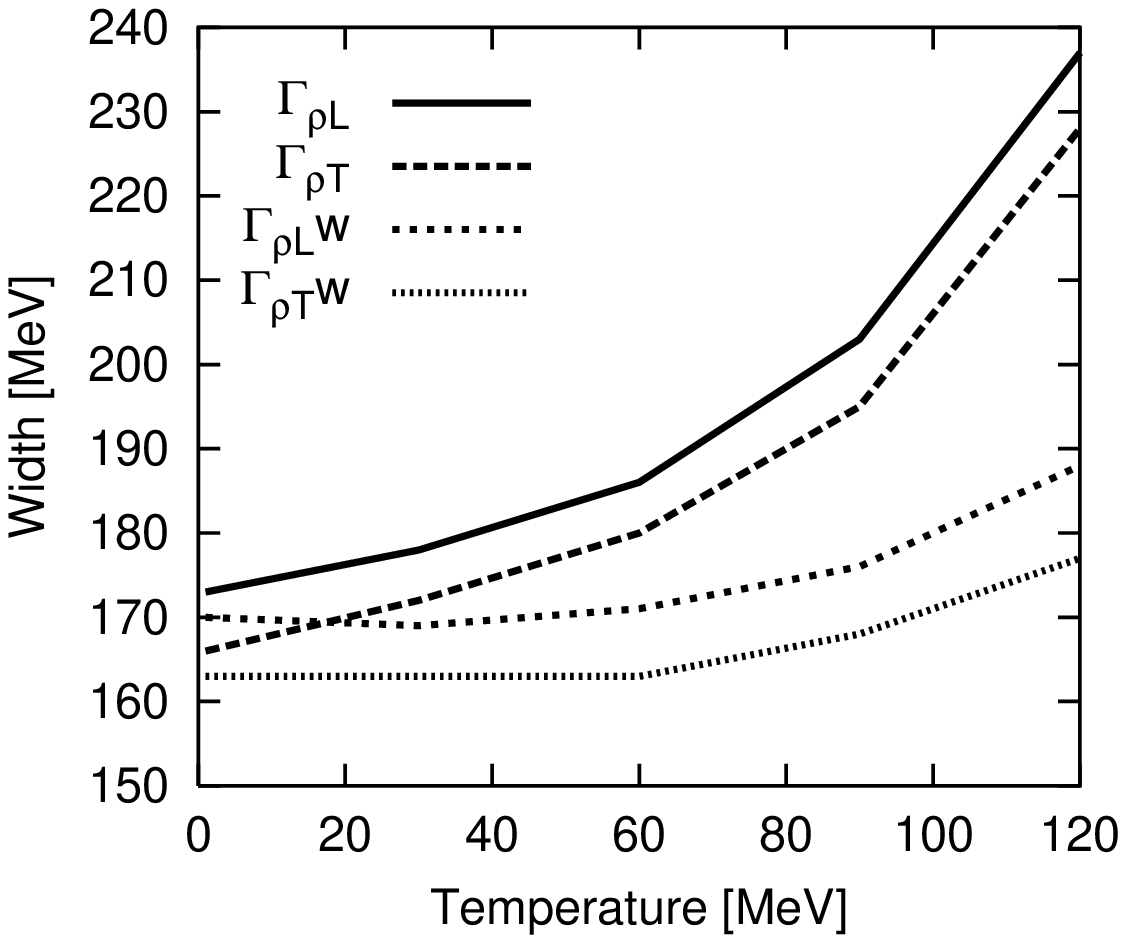,height=6.6 cm,width=5.65 cm,angle=-90}
         \caption{$\Gamma_\rho$  depending on the temperature. $\Gamma$ w marks the process with only $\rho\rightarrow\pi\pi$.}
         \end{minipage}\hfill
        \begin{minipage}[t]{.48\linewidth}
         \epsfig{file=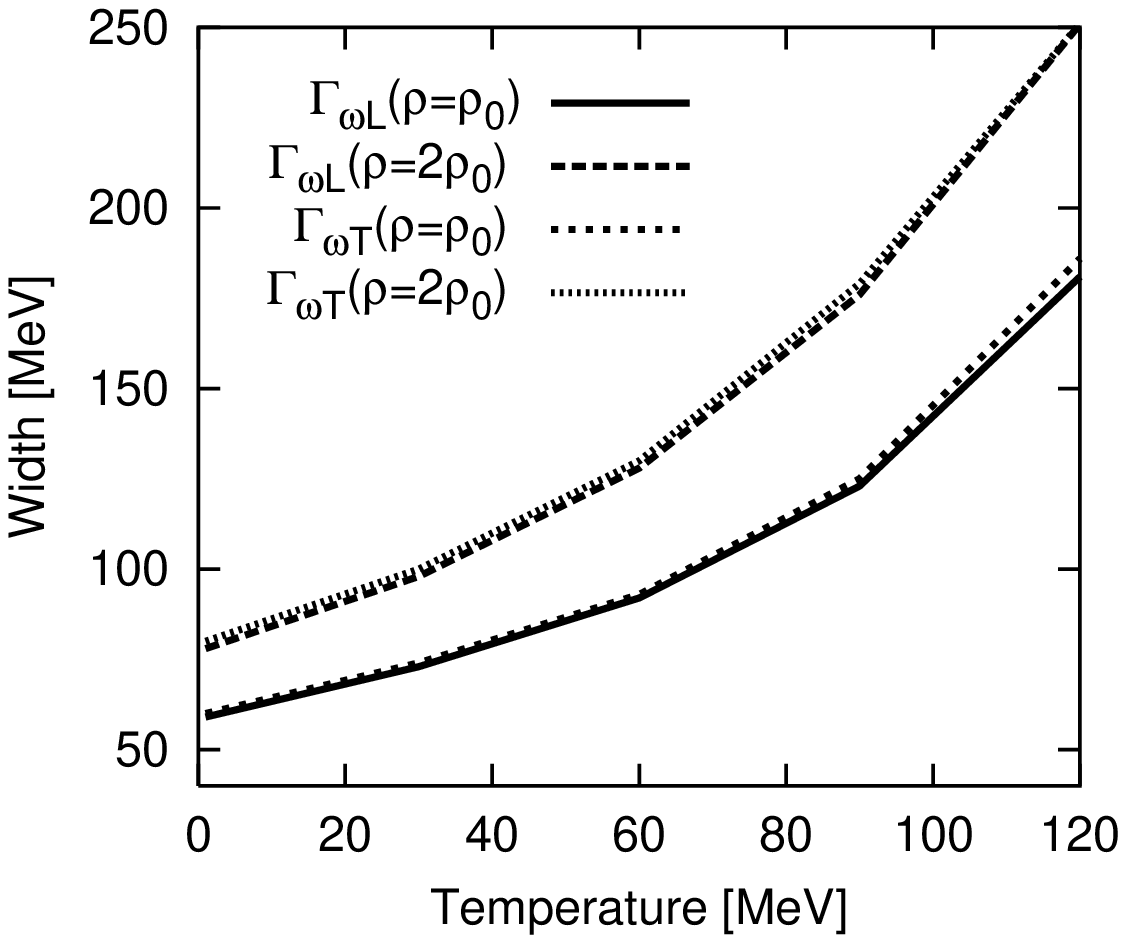,height=6.6 cm,width=5.65 cm,angle=-90}
         \caption{$\Gamma_\omega$  depending on the
           temperature.\label{Breite-Rho-mit}} 
         \end{minipage}
\end{figure}

We restrict the discussion of the dependence of the vector-meson
properties on density and temperature to the case where the baryons
and the vector-mesons retain their vacuum masses and use a cut-off
$\Lambda=440$ MeV (\ref{formfactor}) at the pion-nucleon- and
pion-nucleon-$\Delta$-vertices (\ref{Baryon-lag}). The results for the
damping width at resonance condition, i.e. at the vacuum masses, are
summarised in Figs.  \ref{Breite-Rho-Dichte-mit} to
\ref{Breite-Rho-mit}.

Both vector-mesons show a strong non-linear dependence on the baryon
density $\rho$ induced by the selfconsistent treatment. For the
$\rho$-meson the relative changes are less dramatic due to the large
vacuum width of 150 MeV from the $\pi\pi$-annihilation channel. The
temperature effects are different. Here, with increasing temperature,
the $\rho$-meson width stays nearly constant at nuclear saturation
density $\rho_0$.  The main contribution to the increase of the total
width comes from the processes $\rho\rightarrow\omega\pi$ and
$\rho\pi\rightarrow\omega$.  On the other hand the $\omega$-meson is
seen to have a strong dependence on temperature inducing a width
change of up to a factor of 3 in the temperature range from T=0 MeV to
T=120 MeV.  For both mesons the temperature effects are due to an
enhanced scattering with the pions in the medium\footnote{This has
  already been pointed out by Roy et al. \cite{Roy} who also indicated
  that the process $\omega\rightarrow\rho\pi$ might be increased
  in-matter due to mass shifts.}. In comparison to the pertubative
calculations by Schneider and Weise \cite{Schneider1} the
selfconsistent treatment produces an even broader $\omega$ spectral
function, due to the appearance of the low energy, space-like, pion
modes. The QCD sum rules approach of Klingl and Weise \cite{Klingl3}
and also the low-density expansion scheme of Lutz et.al.  \cite{Lutz3}
produced $\omega$-meson widths of about 40 MeV at $T=0$ and $\rho
=\rho_0$, which is nearly 2/3 of ours.  The slight deviations to our
results can be understood by the change of the vector-meson masses in
medium in their models as discussed in chapter \ref{mass-red}.
Compared to the $\sigma$-$\omega$-mixing model of Saito et al.
\cite{Saito} we cannot account for level-level repulsion in our model,
as we do not include the real parts of the vector-meson selfenergies.
         
Altogether the $\omega$-meson width sensitively depends on the
properties on the in-medium pion cloud. An increase of the
cut-off scale $\Lambda$ of the $\pi NN$-formfactor (\ref{formfactor})
for instance would significantly enhance the RPA pion-modes and induce 
a further
increase of the $\omega$-meson width. Once the $\omega$-meson width
increases beyond 50 MeV, it loses its prominent peak-structure and its
component can no longer be easily resolved in the dilepton spectrum
due to the dominance of the $\rho$ contribution.

%%%%%%%%%%%%%%%%%%%%%%%%%%%%%%%%%%%%%%%%%%%%%%%%%%%%%%%%%%%%
\subsection{Dependence on the vector-meson masses}\label{mass-red}
%%%%%%%%%%%%%%%%
Our description of the vector-meson sector is not as complete as to
permit definite predictions for the mass shifts of the vector-mesons,
i.e. the real parts of the vector-meson selfenergies. This is a
subject of current debate and investigations and requires a more
complete model space within hadronic descriptions \cite{Peters1,Lutz3}
or support from QCD, e.g. in form of sum-rule constraints
\cite{Kaiser,Kaempfer,Hatsuda,Leupold,Jin}. For simplicity the real
parts of the vector-meson selfenergies are put to zero in the results
presented above. On the other hand we can explore the effect of mass
shifts by simply changing the masses by hand. For the following
discussion we concentrate on the damping widths (longitudinal and
transverse) of the vector-mesons at the in-medium mass position
$q^2=m_{\rm{v}}^2$ for a typical momentum of 200 MeV/c given by
\begin{equation}
\Gamma_{\rm{v},L/T}=\left.-\Im
  \Pi^R_{\rm{v},L/T}(q)/q_0\right|_{q^2=m_{\rm{v}}^2}. 
\end{equation}
\begin{figure}[t]
         \begin{minipage}[t]{.48\linewidth}
         \epsfig{file=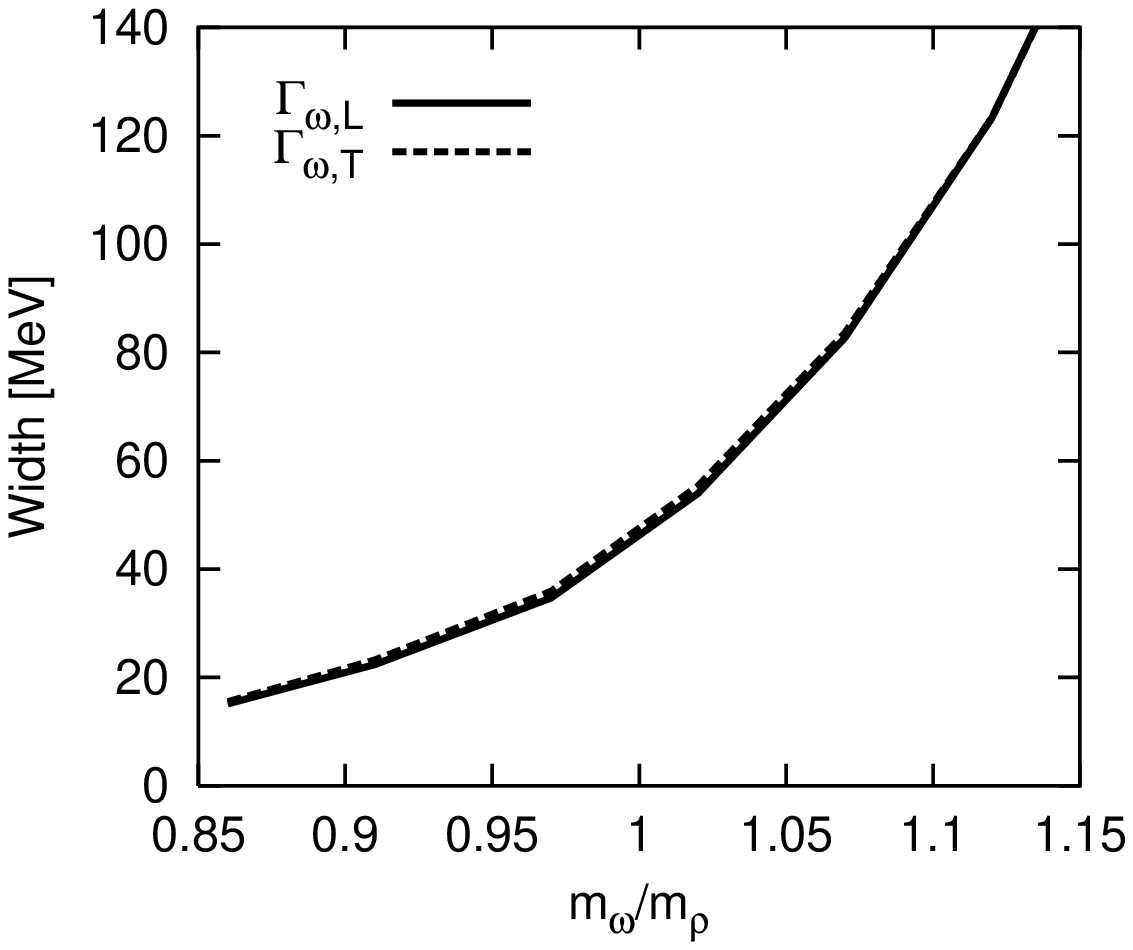,height=6.6
         cm,width=5.6 cm,angle=-90}
         \caption{$\Gamma_\omega$ at T=0 MeV and $\rho=\rho_0$
         depending on the mass ratio $m_\omega/m_\rho$.\label{mass-omega}}
         \end{minipage}\hfill
         \begin{minipage}[t]{.48\linewidth}
         \epsfig{file=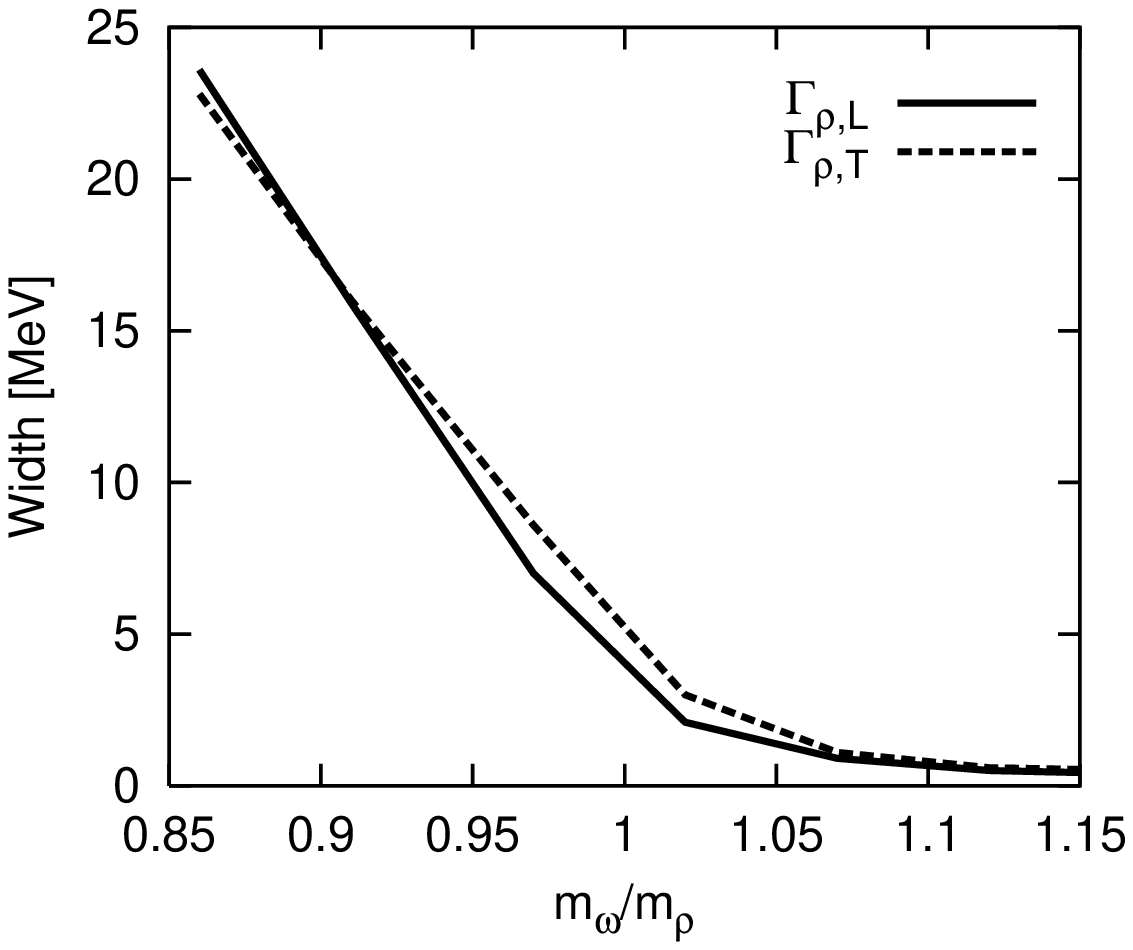,height=6.6
         cm,width=5.6 cm,angle=-90}
         \caption{Partial decay width $\Gamma_{\rho\rightarrow\omega\pi}$ for 
           the process
         $\rho\rightarrow\omega\pi$ at T=0 MeV and $\rho=\rho_0$
         depending on the mass ratio $m_\omega/m_\rho$.\label{mass-rho}}
         \end{minipage}\\
\begin{minipage}[t]{.48\linewidth}
\epsfig{file=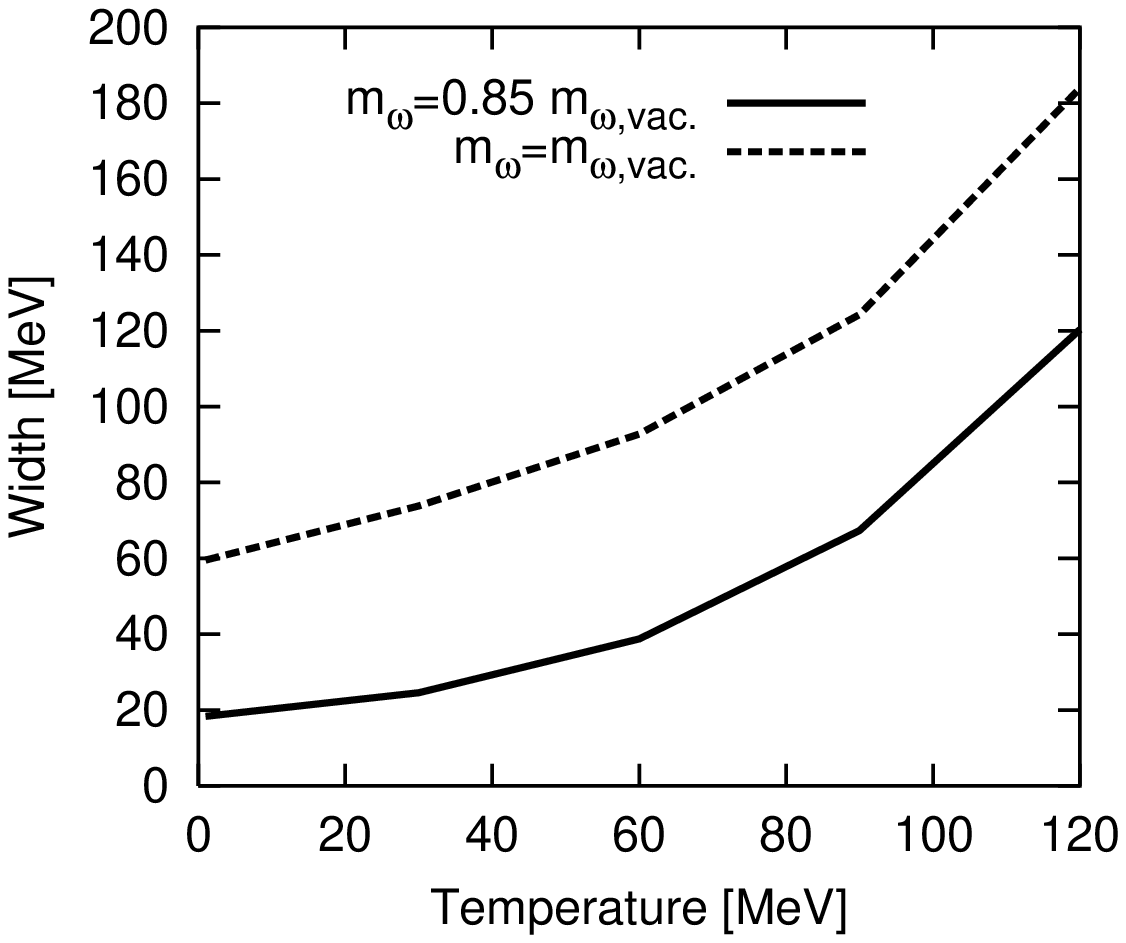,height=6.6 cm,width=6.0 cm,angle=-90}
\caption{Width of the $\omega$-meson with and without reduced mass for normal 
nuclear density and different temperatures.\label{mass-comp}}
\end{minipage}\hfill
\begin{minipage}[t]{.48\linewidth}
\epsfig{file=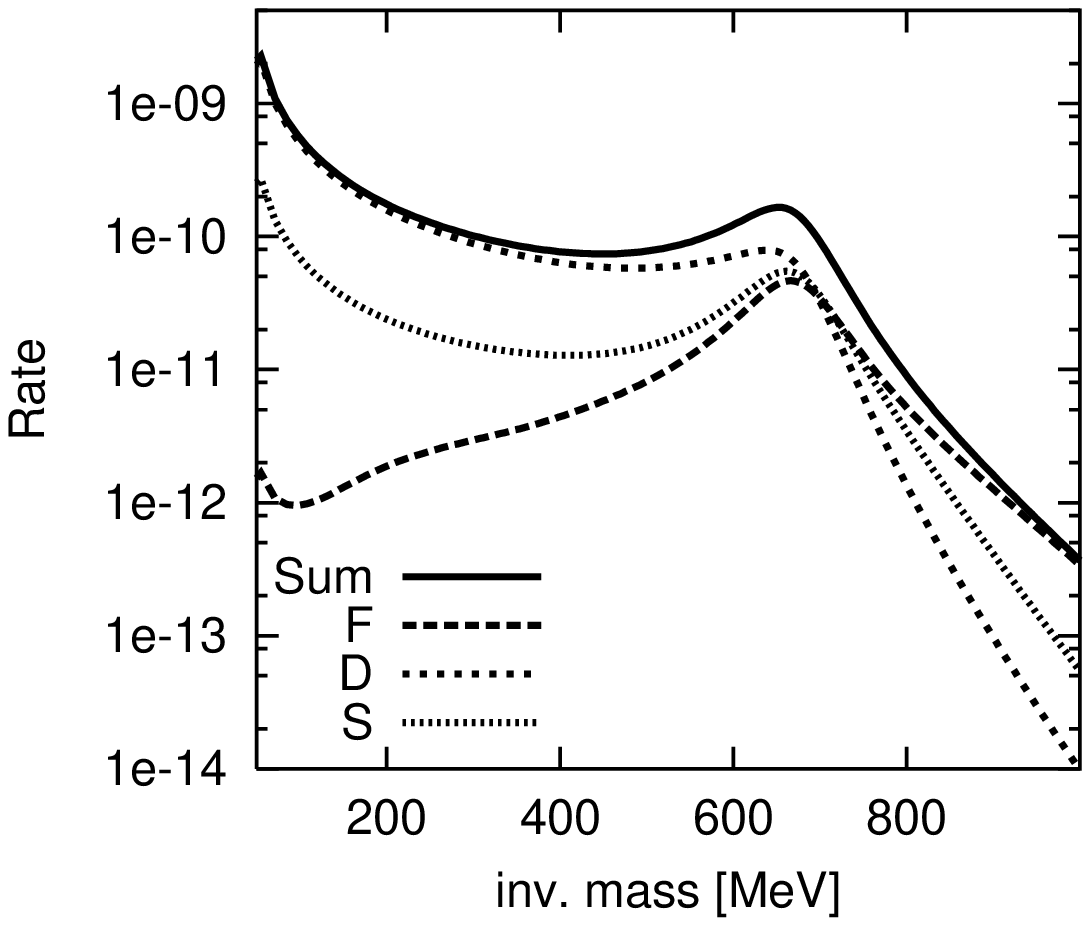,height=6.6 cm,width=6.0 cm,angle=-90}
\caption{$e^+e^-$-spectra resulting from the decay of the $\omega$-meson at
$m_\omega=0.85m_{\omega ,vac.}$, T=120 MeV and $\rho=\rho_0$, For notations
see Fig. \ref{omega-div}.\label{mass-dilepton}}
\end{minipage}
\end{figure} 

A phenomenological lowering of the $\rho$-meson mass, e.g. within the
Brown-Rho scaling scheme \cite{Brown}, has been considered by many
authors. This leads to an enhanced strength in the mass region below
the $\rho$-meson vacuum mass.  From the calculations thus far
the mass dependence of the vector-meson damping widths at $T=0$ and
saturation density can approximately be read off by i.e.  inspecting
Figs. \ref{fig.8} and \ref{fig.12}.  They predict marginal changes for
the $\rho$-meson width. Thus, the entire $\rho$-meson strength
will be shifted, e.g. towards lower masses, with the extra benefit
that the thermal weight factor in (\ref{d R}) enhances the dilepton
yield by about a factor $\exp(\Delta m_{\rho}/ T)$.
The $\omega$-meson shows sensitive changes of its damping width as a
function of its invariant mass. Since the only channel, to which the
$\omega$-meson couples to in our model, is the $\pi$-$\rho$-channel, we
expect $\Gamma_{\omega}$ to depend sensitively only on the mass
difference or ratio between the $\rho$- and $\omega$-meson masses.

The explicit results for the dependence on $m_{\omega}/m_{\rho}$ for
$T=0$ and normal nuclear density can be found in the Figs.
\ref{mass-omega} and \ref{mass-rho} for the $\omega$-meson and for the
partial width $\Gamma_{\rho\rightarrow\omega\pi}$ of the $\rho$-meson,
respectively.  Due to the dominance of the $\pi\pi$-annihilation
channel, the latter has minor impact on the total $\rho$-meson width.
The displayed dependence follows expectations from phase-space
arguments. Our results compare favourably with the calculations of
Klingl and Weise \cite{Klingl3} and Lutz et.al.  \cite{Lutz3}, who
used QCD sum rules or the low density theorem for the determination of
the mass shifts. Although in the Klingl and Weise result
\cite{Klingl3} the mass of the $\omega$-meson is lowered by 15\%
relative to the $\rho$-meson mass, which would infer a much smaller
width, one has to take into account that in those calculations a much
broader $\rho$-meson ($\Gamma_{\rho}\approx 300$ MeV) was used
inducing a compensating increase of the $\omega$-width.
We also studied the vector-meson mass dependence at finite
temperature. In Fig.  \ref{mass-comp} we compare the width of the
$\omega$-meson with a mass reduced by 15\% to the width without such a
reduction\footnote{We show the mean value of longitudinal and
  transverse widths $\Gamma_L$ and
  $\Gamma_T$.}.  We find that the overall offset is nearly the same
for all temperatures.  According to this, the relative effect decreases
for higher temperatures. This can be understood because kinematical
effects, like mass shifts, are of less importance for broad spectral
functions. In Fig. \ref{mass-dilepton}, we show the effect of this mass
shift on the dilepton spectra. We observe that the peak structure
becomes more prominent because of the smaller width of the
$\omega$-meson and that the total rate is enhanced. This enhancement
is due to the thermal weight which increases exponentially towards lower
temperature. 

%%%%%%%%%%%%%%%%%%%%%%%%%%%%%%%%%%%%%%%%%%%%%%%%%%%%%%%%%%%%

\section{Conclusions}
We have investigated the in-medium effects on the light vector-mesons due
to the modification of the pion modes in nuclear matter at finite
density and temperature. To isolate this effect only the coupling of
the vector-mesons to the pion modes has been considered so far. The
direct coupling to baryonic currents along the lines of refs.
\cite{Peters1,Post1,Lutz3} within a selfconsistent scheme will be the
subject of a forthcoming study.

Within a selfconsistent Dyson resummation scheme, the nucleon and
$\Delta$(1232)-resonance were included as the main degrees of freedom
in the baryon sector.  Besides the direct $\pi NN$ and $\pi N\Delta$
couplings, accounted for in selfconsistent selfenergies up to one-loop
order, also the short-range RPA correlations of Migdal type
\cite{migdal} were included in order to get a realistic behaviour of
the low energy pion modes.  While for the meson-baryon couplings a
non-relativistic approximation was employed, the kinematics of all
particles was treated relativistically.  The so obtained in-medium
spectral functions of the pion provide the source for the vector-meson
selfenergies through the $\pi\pi\rho$ and $\pi\rho\omega$ coupling
vertices.  While the former already induces the strong damping width
of the $\rho$-meson in vacuum, the latter becomes essential for the
strong vector-meson ``mixing'' at finite densities and temperatures.
Consequences for the resulting in-matter dilepton spectra from the
electromagnetic decay of the vector-mesons have been discussed.
%%%%%%%%%%%%%%%%%%%%%%%%%%%%%%%%%%%%%%%%%%%%%%%%%%%%%%%%%%%

The polarisation-tensor for the vector-mesons has to be four-transversal
such that no unphysical degrees of freedom are propagated. This was
achieved by a projector technique (see Eq. (\ref{projektion})
ff.) which provides the three-longitudinal and three-transversal parts,
respectively. These two parts are identical in the vacuum due to
Lorentz invariance, while they may differ in the medium.

As an important result of our investigations we find a strong mixing
of both vector-meson modes in-matter. It is induced by the
low-lying space-like pion modes which essentially arise from the
coupling to particle-hole excitations.  As a consequence, the
$\omega$-meson damping width drastically broadens both, with
increasing baryon density and temperature reaching values of 100 MeV
or even above. 

The quantitative results on the in-medium properties of the vector
mesons still depend on the model assumptions and parameters for the
in-medium pion physics such as the Migdal parameters and the formfactors
for the $\pi NN$ and $\pi N\Delta$ vertices. Despite more than two
decades of work in this field, the in-medium properties of the
pion are not yet conclusively settled. Thus, further improvements,
especially concerning the $NN^{-1}$-component, are required for a
quantitative understanding of the in-medium broadening of the light
vector-meson spectral functions, especially of the $\omega$-meson.
Here other techniques like the recent calculations from Lutz and Korpa
\cite{Lutz1} are to be mentioned which may offer alternative and
quantitative strategies to understand the spectral function of the
pion and thus provide a reliable basis for the calculations of the
vector-mesons properties.  As our approach qualitatively incorporates
most of the relevant contributions to the pion spectral function, all
qualitative features should already be visible in our results.  Even
if the situation at zero temperature and saturation density may eventually
become quantitatively settled, the extrapolation of the effective
interactions to finite temperatures may requires further adaptions.

The broad $\omega$-meson distributions, as found in our calculations,
will make it difficult to isolate this in-medium component in dilepton
spectra. It will be detected in competition with the much stronger
electromagnetic decay rates of the $\rho$-meson.  Therefore
complementary experiments which directly isolate the
$\omega$-component e.g. partly through hadronic channels, such as in
the recent TAPS-experiments \cite{Messch,Metag}, are vital to clarify
the situation.  In nuclear collision experiments, one further has to
face the fact that contributions from final-state interaction always
arise.  Such asymptotic state vector-mesons show the signature of the
corresponding vacuum spectral functions and thus may hide the here
addressed in-matter components.\\[-10mm]
\section*{Acknowledgement}
The authors acknowledge fruitful discussions with C. Greiner, B.
Friman, C.L. Korpa, S. Leupold, M.F.M. Lutz, D. Voskresensky and J.
Wambach at various stages of this work.

\end{document}